\documentclass[useAMS,usenatbib]{mnras} % for a referee version
\usepackage{graphicx}
\usepackage{color}
\usepackage[usenames,dvipsnames]{xcolor}
\usepackage{mathrsfs}
\usepackage{lscape}
\usepackage{txfonts}
\usepackage{epstopdf}

\definecolor{bdf}{rgb}{0.19, 0.55, 0.91}

\title[Evolution of CNO isotopes]{The evolution of CNO isotopes: a new window 
  on cosmic star-formation history and the stellar IMF in the age of ALMA}

\author[D.~Romano et al.]{D.~Romano,$^{\! 1}$\thanks{E-mail: 
    donatella.romano@oabo.inaf.it} F.~Matteucci,$^{\!2, 3, 4}$ 
  Z.-Y.~Zhang,$^{\!5, 6}$ P.\,P.~Papadopoulos$^{\!7, 6, 8}$ and 
  R.\,J.~Ivison$^{\!5, 6}$\\
  $^{1}$ INAF, Osservatorio Astronomico di Bologna, Via Ranzani 1, I-40127 
  Bologna, Italy\\
  $^{2}$ Dipartimento di Fisica, Sezione di Astronomia, Universit{\`a} di 
  Trieste, Via Tiepolo 11, I-34131 Trieste, Italy\\
  $^{3}$ INAF, Osservatorio Astronomico di Trieste, Via Tiepolo 11, I-34131 
  Trieste, Italy\\
  $^{4}$ INFN, Sezione di Trieste, Via Valerio 2, I-34127 Trieste, Italy\\
  $^{5}$ Institute for Astronomy, University of Edinburgh, Royal Observatory, 
  Blackford Hill, Edinburgh EH9 3HJ, UK\\
  $^{6}$ European Southern Observatory, Karl-Schwarzschild-Str.~2, D-85748 
  Garching bei M\"unchen, Germany\\
  $^{7}$ School of Physics and Astronomy, Cardiff University, Queen's 
  Buildings, The Parade, Cardiff CF24 3AA, UK\\
  $^{8}$ Department of Physics, Section of Astrophysics, Astronomy and 
  Mechanics, Aristotle University of Thessaloniki, Thessaloniki 54124, Greece}

\begin{document}

\date{Accepted 2017. Received 15 March 2017; in original form 23 December 2016}

\pagerange{\pageref{firstpage}--\pageref{lastpage}} \pubyear{2017}

\maketitle

\label{firstpage}

%%%%%%%%%%%%%%%%%%%%%%%%%%%%%%%%%%%%%%%%%%%%%%%%

\begin{abstract}
We use state-of-the-art chemical models to track the cosmic evolution of the 
CNO isotopes in the interstellar medium (ISM) of galaxies, yielding powerful 
constraints on their stellar initial mass function (IMF). We re-assess the 
relative roles of massive stars, asymptotic giant branch (AGB) stars and novae 
in the production of rare isotopes such as $^{13}$C, $^{15}$N, $^{17}$O and 
$^{18}$O, along with $^{12}$C, $^{14}$N and $^{16}$O. The CNO isotope yields of 
super-AGB stars, novae and fast-rotating massive stars are included. Having 
reproduced the available isotope enrichment data in the solar neighbourhood, 
and across the Galaxy, and having assessed the sensitivity of our models to the 
remaining uncertainties, e.g.\ nova yields and star-formation history, we show 
that we can meaningfully constrain the stellar IMF in galaxies using C, O and N 
isotope abundance ratios. In starburst galaxies, where data for multiple 
isotopologue lines are available, we find compelling new evidence for a 
top-heavy stellar IMF, with profound implications for their star-formation 
rates and efficiencies, perhaps also their stellar masses. Neither chemical 
fractionation nor selective photodissociation can significantly perturb 
globally-averaged isotopologue abundance ratios away from the corresponding 
isotope ones, as both these processes will typically affect only small mass 
fractions of molecular clouds in galaxies. Thus the Atacama Large Millimetre 
Array now stands ready to probe the stellar IMF, and even the ages of specific 
starburst events  in star-forming galaxies across cosmic time unaffected by the 
dust obscuration effects that plague optical/near-infrared studies.
\end{abstract}


%%%%%%%%%%%%%%%%%%%%%%%%%%%%%%%%%%%%%%%%%%%%%%%%

\section{Introduction}
\label{sec:intro}

Measurements of isotope abundances and abundance ratios allow us to perform key 
tests of mixing mechanisms inside stars, and provide powerful diagnostics of 
chemical enrichment in galaxies across cosmic time. Some carry valuable 
information about the state of the early Universe, which can be used to probe 
fundamental physics.  Determinations of the $^6$Li/$^7$Li ratios in metal-poor 
stars \citep{2007A&A...473L..37C, 2009A&A...504..213G, 2013A&A...554A..96L}, 
for instance, constrain Big Bang nucleosynthesis theories 
\citep{2016RvMP...88a5004C} and the degree of $^7$Li depletion possible in 
Galactic halo stars \citep{2010IAUS..268..201S}. The CNO isotope ratios 
measured in presolar grains \citep{2004ARA&A..42...39C, 2014mcp..tog..181Z} and 
in the atmospheres of evolved stars \citep[][and references 
therein]{2000A&A...354..169G, 2009A&A...502..267S} can be compared to 
expectations from stellar evolution and nucleosynthesis models to shed light on 
the nature of non-standard mixing processes acting on the giant branches 
\citep[see][among others]{1994A&A...282..811C, 1995ApJ...447L..37W,
2007A&A...467L..15C, 2009PASA...26..161P}. Peculiar isotopic features in C-rich 
grains can be explained by considering the effects of hydrogen ingestion into 
the helium shell of core-collapse supernovae before the shock hits the outer 
layers \citep{2015ApJ...808L..43P}. Beginning with \cite{1968ApJ...154..185B}, 
magnesium isotope abundances have been measured in stars in order to provide 
insights into the production sites of the two minor, neutron-rich isotopes, 
$^{25}$Mg and $^{26}$Mg. Last but not least, measurements of barium \citep[][and 
references therein]{2015A&A...579A..94G}, europium, samarium and neodymium 
isotope ratios in stars \citep{2008ApJ...675..723R} allow us to establish 
whether the dominant formation process is slow or rapid neutron capture by 
heavy seed nuclei, thus contributing significantly to our understanding of how 
neutron-capture elements are created in galaxies.

In principle, large samples of stars with precise chemical abundances over the 
full range of plausible metallicities should provide a complete fossil record 
of the history of chemical enrichment for their host galaxies \citep[provided 
the data are corrected for stellar evolutionary effects when needed; 
see][]{2014ApJ...797...21P}. Alongside accurate determinations of stellar ages, 
distances and kinematics, abundance data can meaningfully constrain models of 
galaxy formation and evolution \citep{2002ARA&A..40..487F,2012ARA&A..50..251I}. 
However, when it comes to the determination of isotopic abundance ratios in 
stars, very high-resolution, high signal-to-noise spectroscopic data are 
invariably needed, which severely limits current observational studies. It is 
then useful to turn to measures of the gas-phase isotopic abundances in 
interstellar clouds, which enables us to look significantly further afield. On 
the other hand, each observation provides merely a snapshot in time, and 
information about the temporal sequence of events that led to the observed 
configuration is thus missing. Based on these diagnostics, however, crucial 
information on the present-day gradients of $^{12}$C/$^{13}$C, $^{14}$N/$^{15}$N, 
$^{16}$O/$^{18}$O and $^{18}$O/$^{17}$O across the Milky Way disc can be obtained 
\citep{1994ARA&A..32..191W}.

Indeed, in the era of the Atacama Large Millimetre Array (ALMA) it has become 
possible to peer back to the epoch when gas-rich galaxies dominated the 
Universe, at $z>1$, which should lead to a better understanding of the 
evolutionary links between these galaxies and their descendents. Exploiting 
isotopologue line intensity ratios as probes of the corresponding elemental 
isotope ratios will open a new window onto the isotope enrichment history of 
the Universe across cosmic epoch.  Such observations were already plentiful in 
the local Universe, even before ALMA \citep[e.g.][]{1991A&A...249...31S,
1992A&A...264...55C,1993A&A...274..730H,1995A&A...300..369A,1996ApJ...465..173P,
2001ApJS..135..183P,2009ApJ...692.1432G}, and now extend out to $z \sim 3$ 
\citep[][Zhang et al., in preparation]{2010A&A...516A.111H,
2014ApJ...785..149S}. Even more importantly, isotopologue line ratios are the 
\emph{only} probe of isotope ratios that are unaffected by dust extinction, as 
isotopologue lines (e.g.\ $^{12}$CO, $^{13}$CO, C$^{18}$O, H$^{13}$CN, 
HC$^{15}$N rotational transitions) are found in the millimetre/submillimetre 
(submm) regime. Besides the obvious advantage of a method insensitive to dust 
extinction, it is in the highly dust-enshrouded star-forming environments of 
compact starbursts \citep[e.g.][]{2015ApJ...810..133I,2016arXiv161103084S} 
where {\it very different average initial conditions of star formation} 
\citep{2010ApJ...720..226P} can lead to a different stellar initial mass 
function \citep[IMF, see][]{2011MNRAS.414.1705P}. Yet it is exactly in these 
dust-obscured galaxies where the stellar IMF is inaccessible via standard 
methods (e.g.\ star counts) where the different isotope enrichment of the 
interstellar medium (ISM) would provide the best evidence for a different 
stellar IMF. In short, in the dust-obscured environments of star-forming 
galaxies, {\it isotope abundance ratios provide the next best constraint on the 
prevailing stellar IMF, other than starlight itself.} This prevailing IMF is 
important to our understanding of galaxy evolution since it is a key ingredient 
of the recipes used to determine the instantaneous star-formation rate, for 
example, where extrapolations are made from the number of massive stars to the 
total mass of stars. The evolution of galaxies is often explored in the 
observational framework of the so-called `main sequence' 
\citep[e.g.][]{2011A&A...533A.119E}, plotted usually as star-formation rate 
versus stellar mass. Where both measurements are sensitive to the IMF, the 
implications may be profound.

In past years many theoretical efforts \citep[including][]{1975A&A....43...71A,
1978ApJ...223..557D,1982ApJ...254..699T,1991A&A...247L..37M,1992A&ARv...4....1W,
1995ApJS...98..617T,1996A&A...309..760P,2003PASA...20..340F,2003MNRAS.342..185R,
2008A&A...479L...9C,2008MNRAS.390.1710H,2011MNRAS.414.3231K} have been devoted 
to the evolution of isotope ratios in the Galaxy. \cite{2008MNRAS.390.1710H} 
also discuss observations of carbon and sulphur isotopes in a spiral galaxy and 
a damped Ly$\alpha$ system at $z \sim 1$. Although there is no doubt that 
progress has been made, some questions remain unanswered. For example, we know 
that $^{12}$C is produced as a primary element in stars \citep[i.e.\ starting 
from the original H and He;][]{1957RvMP...29..547B} and that the relative 
contributions from low- and intermediate-mass stars and from massive stars 
change with time and position within the Galaxy 
\citep[e.g.][]{2004A&A...414..931A}; the actual proportions are still debated, 
however.

Regarding $^{14}$N, it has been acknowledged for a long time that significant 
primary nitrogen production is needed to explain observations of 
low-metallicity systems \citep[e.g.][]{1978MNRAS.185P..77E,
1986MNRAS.221..911M}. The physical process responsible for this production 
remained elusive for decades, until very low-metallicity, fast-rotating massive 
star models proved able to convert part of the freshly made $^{12}$C into 
primary $^{14}$N (and $^{13}$C), quite efficiently, because of rotation-induced 
mixing between the convective hydrogen shell and the helium core 
\citep{2002A&A...381L..25M, 2008A&A...479L...9C}. However, chemical evolution 
models that adopt $^{14}$N yields from fast-rotating stellar models still 
predict a much lower nitrogen abundance in the ISM than observed in the 
metallicity range $-2.5<{\rm [Fe/H]}< -1.5$ \citep{2006A&A...449L..27C,
2010A&A...522A..32R}. In intermediate-mass stars climbing the asymptotic giant 
branch (AGB), proton-capture nucleosynthesis leads to primary $^{14}$N (and 
$^{13}$C) production if the base of the convective envelope becomes hot enough 
\citep[a process referred to as hot bottom burning; 
see][]{1975ApJ...196..525I}. Moreover, a large fraction of $^{14}$N is 
synthesised as a secondary element, at the expense of $^{12}$C, through the CN 
cycle in stars of all masses and --at a much slower rate-- at the expense of 
$^{16}$O in the ON cycle. Adding these contributions, however, does not help to 
solve the aforementioned problem.

Another outstanding issue is that of $^{15}$N evolution. Novae have been 
identified as $^{15}$N polluters on a Galactic scale 
\citep{1978ApJ...223..557D, 1991A&A...247L..37M, 2003MNRAS.342..185R}, but 
observations of low $^{14}$N/$^{15}$N at high redshift 
\citep{2006A&A...458..417M} suggest an additional production channel, possibly 
hydrogen ingestion in the helium shell of massive stars 
\citep{2015ApJ...808L..43P}.

The synthesis of $^{16}$O in stars, on the other hand, is well understood, both 
qualitatively and quantitatively, and chemical evolution models can fit the 
relevant data, regardless of the specific set of yields adopted \citep[see 
figure~8 of][]{2010A&A...522A..32R}. The less abundant, neutron-rich oxygen 
isotopes, $^{17}$O and $^{18}$O, are made as secondary elements by the CNO cycle 
during hydrogen burning and by $\alpha$-captures on $^{14}$N during helium 
burning, respectively. While intermediate-mass stars, massive stars and novae 
all contribute significantly to $^{17}$O production, the situation for $^{18}$O 
is less clear; it seems, in fact, that intermediate-mass stars destroy $^{18}$O 
rather than produce it.

In this work, we focus on the evolution of the CNO isotopes in the ISM of 
galaxies --those associated with the most abundant molecule after $\rm H_2$, 
namely carbon monoxide (CO). First, we re-assess the relative roles of massive 
stars, AGB stars and novae in the production of the rare isotopes $^{13}$C, 
$^{15}$N, $^{17}$O and $^{18}$O on a Galactic scale, in the light of newly 
published stellar yields and more recent isotopic abundance determinations in 
the Milky Way. Second, we discuss how new measurements of $^{12}$C/$^{13}$C and 
$^{16}$O/$^{18}$O ratios in starburst galaxies can constrain their stellar 
IMF. We stress that these constraints on the stellar IMF {\it are 
ultimately set by the stellar physics underlying the chemical evolution 
model,} and as such are extremely powerful, regardless of remaining 
uncertainties. In this work we examine these uncertainties, focusing on the 
assumed stellar yields, star-formation histories and the astrochemical effects 
on molecules, such as fractionation and selective photodissociation. These two 
effects, if widespread, could in principle prevent us from directly deducing 
isotope abundance ratios from the corresponding isotopologues. In the age 
of ALMA, numerous isotopologue line ratios can now be measured for the 
molecular gas of dust-obscured galaxies, where there is no hope of measuring 
the stellar IMF directly via star counts or integrated starlight, leaving these 
ratios as the next best thing for constraining the IMF 
\citep{2014ApJ...788..153P}.

The paper is organised as follows: in \S\ref{sec:data}, we review the available 
data. In \S\ref{sec:mod}, we briefly describe the chemical evolution models. In 
\S\ref{sec:res}, we present the model results and compare them to data for both 
the Milky Way (\S\ref{sec:MWres}) and other galaxies (\S\ref{sec:EXTres}), with 
special emphasis on intense starburst galaxies. We discuss our findings and 
conclusions in \S\ref{sec:end}.

\section{Observational data}
\label{sec:data}

%%%%%%%%%%%%%%%%%%%%%%%%%%%%%%%%%%%%%%%%%%%%%%%% 
%TWO COLUMN TABLE

\begin{table*}\setlength{\tabcolsep}{3pt}
\caption{$^{12}$CO/$^{13}$CO, $^{12}$CO/C$^{18}$O and
       $^{13}$CO/C$^{18}$O abundance ratios for external galaxies, compiled from the literature.}
\begin{tabular}{@{}lllccccll@{}}
\hline
Name                      & Type   & Redshift    & SFR$^a$                & $^{12}$CO/$^{13}$CO  & $^{12}$CO/C$^{18}$O  & $^{13}$CO/C$^{18}$O          & Method                          & References$^b$\\
                          &        &             & (M$_\odot$\,yr$^{-1}$) &                      &                      &                              &                                 &               \\
       \hline
SPT stacking$^c$          & SMG    & 3.0         & 500--2,000             & 100--200             &  $>100$--200         & $>1$ ($3\sigma$)             & LVG modeling                    & 1\\
Cloverleaf                & QSO    & 2.5579      & 1,000                  & 300--10,000          &  --                  & --                           & LVG modeling                    & 2, 3\\
MA2.53                    & DLA    & 2.525       & --                     & $>$ 40               &  --                  & --                           & Optical absorption              & 4\\
Eyelash$^d$               & SMG    & 2.3         & 400                    & 100$^e$              &  100                 & 0.8                          & LVG modeling                    & 5\\
MA1.15                    & DLA    & 1.15        & --                     & $>$ 53 ($3\sigma$)   &  --                  & --                           & Optical absorption              & 6\\
MA0.89                    & Spiral & 0.89        & --                     & $27\pm 2$            &  $52\pm 4$           & 1.9                          & Absorption + LVG modeling       & 7\\
MA0.68                    & Spiral & 0.68        & --                     & $38\pm 5$            &  $\sim 80$           & $\sim$2                      & Absorption + LVG modeling       & 8\\
Mrk\,231                  & ULIRG  & 0.042170    & 100                    & 100                  &  100                 & 1.0                          & CN modeling                     & 9\\
NGC\,6240                 & ULIRG  & 0.0245      & 100                    & 300--500             &  --                  & 1.6                          & LVG modeling                    & 10, 11\\
Arp\,193                  & ULIRG  & 0.023299    & 100                    & $\sim 150$           &  --                  & --                           & LVG modeling                    & 10\\
VV\,114                   & LIRG   & 0.020067    & 48                     & 229                  &  --                  & --                           & LVG modeling                    & 12\\
Arp\,220                  & ULIRG  & 0.018126    & 220                    & --                   &  70--130             & 1.0                          & OH and $\rm H_2O$ absorption    & 13\\
Arp\,220                  & ULIRG  & 0.018126    & 220                    & --                   &  $>80$--100          & 1.0                          & Line ratio limits               & 14\\
NGC\,1614                 & LIRG   & 0.015938    & 41                     & 130                  &  --                  & $>$6.6                       & LVG modeling                    & 15\\
LMC                       & Dwarf  & 0.000927    & one clump              & 49                   &  2000                & 27$\pm$9                     & LVG modeling                    & 16, 17\\
NGC\,253                  & LIRG   & 0.000811    & 2.8                    & $>56$                &  $145\pm 36$         & 2.6                          & Line ratio limits + CN modeling & 18, 9,  19\\
M\,82                     & LIRG   & 0.000677    & 4.6                    & $>138$               &  $>350$              & 2.2-3.7                      & Line ratio limits + CN modeling & 18, 20, 21\\
NGC\,1068                 & LIRG   & 0.003793    &  25                    & --                   &  --                  & 3.33                         &                                 & 22\\
NGC\,1614                 & LIRG   & 0.015938    & 55                     & $>$36.5              &  $>80$               & $>2.2$                       &                                 & 23\\
NGC\,4945                 & LIRG   & 0.001878    & $<7.8$                 & $>$17                &  $>61$               & 3.6                          &                                 & 24\\
Cen\,A                    & Radio  & 0.001825    & 0.16                   & 17                   &  $>56$               & $>3$                         & Absorption + modeling           & 25, 26\\
NGC\,2903                 & Normal & 0.001834    & 2.7                    & $>10$                & --                   & $>7$ ($3\sigma$)             &                                 & 27\\
IC\,860                   & LIRG   & 0.011164    & 25                     & $>20$                & $>20$                & 1                            &                                 & 28\\
NGC\,3079                 & Normal & 0.003723    & 9.3                    & $>17$                & $>91.5$              & 5.4                          &                                 & 28\\
NGC\,4194 	              & LIRG   & 0.008342    & 20                     & $>18.7$              & $>47$                & $>2.5$                       &                                 & 28\\
NGC\,7469 	              & LIRG   & 0.016317    & 67                     & $>21$                & $>142$               & 6.8                          &                                 & 28\\
NGC\,7771 	              & LIRG   & 0.014267    & 38                     & $>14$                & $>65$                & 4.7                          &                                 & 28\\
NGC\,660 		          & Normal & 0.002835    & 5.3                    & $>17$                & $>46$                & 6.9                          &                                 & 28\\
NGC\,3556                 & Normal & 0.002332    & 4                      & $>12.5$              & $>161$               & 12.9                         &                                 & 28\\
NGC\,7674                 & LIRG   & 0.028924    & 54                     & $>14.5$              & $>40$                & 2.8                          &                                 & 28\\
UGC\,2866                 & Radio  & 0.004110    & 8                      & $>21$                & $>120$               & 5.8                          &                                 & 28\\
Circinus                  & Normal & 0.001448    & 8                      & $>13$                & $>54$                & 5.1                          &                                 & 29, 30, 31\\
IC\,10                    & Normal &$-$0.001161  & 0.2                    & $>7$                 & $>100$ ($3\sigma$)   & $>15$                        &                                 & 32, 33\\
IC\,342                   & Normal & 0.000103    & 2.5                    & $>10$                & $>70$                & 3-7                          &                                 & 34\\ 
M\,51                     & Normal & 0.002000    & 4.5                    & $>10$                & $>46$                & 4.5                          &                                 & 35\\
Maffei\,2                 & Normal &$-$0.000057  & 0.26                   & $>10$                & $>43$                & 4.3                          &                                 & 36\\
NGC\,1808                 & Normal & 0.003319    & 8.8                    & $>17$                & $>49$                & 3                            &                                 & 37\\ 
NGC\,3256                 & LIRG   & 0.009354    & 63                     & $>33$                & $>135$               & 4                            &                                 & 37\\ 
NGC\,7552                 & LIRG   & 0.005365    & 18                     & $>14$                & $>35$                & 3                            &                                 & 37\\ 
NGC\,4826                 & Normal & 0.001361    & 0.2                    & $>8$                 & $>21$                & 4                            &                                 & 37\\ 
NGC\,2146                 & LIRG   & 0.002979    & 20                     & $>12.5$              & $>36$                & 3                            &                                 & 37\\ 
NGC\,4418                 & LIRG   & 0.007268    & 20.7                   & --                   & --                   & 8.3                          &                                 & 38\\
IRAS\,04296+2923 starburst& Normal & 0.007062    & 16                     & $>21$                & $>94$                & 3.7                          &                                 & 39\\
IRAS\,04296+2923 CNZ      & Normal & 0.007062    & 16                     & $>16$                & $>45$                & 1.7                          &                                 & 39\\
NGC\,6946                 & Normal & 0.000133    & 2.5                    & $>$ 13.3             & $>21$                & 2.5                          &                                 & 40\\
\hline
\end{tabular}
\label{tab:ext}
\begin{flushleft}
\emph{Notes.} $^a$ The star-formation rates are from standard IMFs. 
$^b$
1:  \cite{2014ApJ...785..149S}; 
2:  \cite{2010A&A...516A.111H}; 
3:  \cite{2007ApJ...661L..25L}; 
4:  \cite{2017A&A...597A..82N};
5:  \cite{2013MNRAS.436.2793D}; 
6:  \cite{2006A&A...447L..21L}; 
7:  \cite{2006A&A...458..417M}; 
8:  \cite{2016A&A...595A..96W}; 
9:  \cite{2014A&A...565A...3H}; 
10: \cite{2014ApJ...788..153P}; 
11: \cite{2004A&A...415..103P}; 
12: \cite{2013ApJ...777..126S}; 
13: \cite{2012A&A...541A...4G}; 
14: \cite{2011A&A...527A..36M}; 
15: \cite{2014ApJ...796L..15S}; 
16: \cite{2009ApJ...690..580W}; 
17: \cite{1998A&A...332..493H};
18: \cite{2010A&A...522A..62M}; 
19: \cite{1999MNRAS.303..157H}; 
20: \cite{2000A&A...358..433M}; 
21: \cite{2011RAA....11..787T}; 
22: \cite{1999ApJ...516..114P}; 
23: \cite{2016A&A...594A..70K}; 
24: \cite{2001A&A...367..457C}; 
25: \cite{2010ApJ...720..666E}; 
26: \cite{2016A&A...586A..45S}; 
27: \cite{2016PASJ...68...89M}; 
28: \cite{2011A&A...528A..30C}; 
29: \cite{2014A&A...568A.122Z}; 
30: \cite{2014MNRAS.445.2378D}; 
31: \cite{2012MNRAS.425.1934F}; 
32: \cite{2016ApJ...829...94N}; 
33: \cite{2010A&A...520A..55Y}; 
34: \cite{2001ApJ...551..687M}; 
35: \cite{2014ApJ...788....4W}; 
36: \cite{2008ApJ...675..281M}; 
37: \cite{1995A&A...300..369A}; 
38: \cite{2015A&A...582A..91C}; 
39: \cite{2014ApJ...795..107M}; 
40: \cite{2004AJ....127.2069M}. 
$^c$Stacking results of the strongly lensed SMGs found by the South Pole 
Telescope (SPT) survey. All galaxies have been shifted to $z=3$. The C$^{18}$O 
lines only have a 3-$\sigma$ upper limit, while the $^{13}$CO lines were 
detected at the $\sim$3-$\sigma$ level. The infrared (IR) luminosity is 
$L_{\mathrm{IR}}=4.2\times 10^{13}$\,L$_\odot$ which translates to a star-formation 
rate, $\rm SFR \simeq 6000$\,M$_\odot$\,yr$^{-1}$. However, these are lensed 
systems, therefore the inferred SFRs have been scaled down using a suitable 
magnification factor. $^d$We adopt the best-fitting results over the whole 
galaxy. The uncertainty is large 
\citep[see figure~7 of][]{2013MNRAS.436.2793D}. $^e$\cite{2011MNRAS.410.1687D} 
report a 3-$\sigma$ limit $^{12}$CO/$^{13}$CO $>$ 60.
\end{flushleft}
\end{table*}

%%%%%%%%%%%%%%%%%%%%%%%%%%%%%%%%%%%%%%%%%%%%%%%%

\subsection{Milky Way galaxy}
\label{sec:MWdata}

The $^{12}$C/$^{13}$C ratio does not display significant variability within the 
Solar System; we adopt the value inferred from CO infrared lines in the solar 
spectrum using 3D convection models 
\citep[91.4$\pm$1.3;][]{2013ApJ...765...46A} as indicative of the ISM 
composition in the solar vicinity 4.5\,Gyr ago. Observations of CO, H$_2$CO 
\citep[taken from the compilation of][]{1994ARA&A..32..191W}, CO$_2$ 
\citep[from][]{2000A&A...353..349B} and CN \citep{2002ApJ...578..211S,
2005ApJ...634.1126M} and their $^{13}$C-bearing isotopologues are used to trace 
the behaviour of the $^{12}$C/$^{13}$C ratio across the disc of the Galaxy at 
the present time.  The gradients derived from CO and CN agree closely, which, 
together with the lack of correlation of the ratios with gas kinetic 
temperature, $T_{\rm kin}$, suggests that {\it the effects of chemical 
fractionation and isotope-selective photodissociation are negligible} 
\citep{2005ApJ...634.1126M}. Indeed there are simple reasons why these two 
astrochemical effects cannot significantly perturb  isotopologue abundance 
ratios from the corresponding isotopic ratios for the bulk of the molecular gas 
reservoir in galaxies, which we briefly outline in \S\ref{sec:fromto}.

The $^{12}$C/$^{13}$C ratios derived from both H$_2$CO and CO$_2$ tend to be 
higher than those derived from other tracers \citep{2000A&A...353..349B}. The 
carbon isotope ratio varies significantly in the local ISM; we take the average 
value of ${\rm ^{12}C/^{13}C} =68\pm 15$ suggested by \cite{2005ApJ...634.1126M} 
as typical of the local ratio. It is worth noting that because of the 
significant heterogeneity in interstellar carbon isotope ratios, it is unclear 
whether the solar value is truly representative of the average local 
$^{12}$C/$^{13}$C ratio 4.5\,Gyr ago.  On top of that, the Sun might have 
migrated to its current position from a birthplace closer to the Galactic 
Centre \citep{1996A&A...314..438W} and its composition could thus reflect 
chemical enrichment occurring on faster time scales. Measurements of 
$^{12}$C/$^{13}$C ratios in statistically significant samples of nearby dwarf 
stars would usefully constrain the models, but they are challenging. In 
brighter, giant stars, on the other hand, mixing may have altered the original 
abundances. In this paper, we use $^{12}$C/$^{13}$C data from 
\cite{2006A&A...455..291S} for a sample of `unmixed' halo giants, i.e.\ stars 
lying mostly on the low red giant branch where the original CNO abundances were 
likely unaltered by mixing processes.

The nitrogen isotope ratio presents extreme variations among different Solar 
System objects \citep[e.g.][]{2015NatGe...8..515F}; we adopt as a proxy for the 
proto-solar nebula the estimate for the bulk Sun from 
\cite{2011Sci...332.1533M}, namely ${\rm ^{14}N/^{15}N} = 441\pm 6$. 
Considerations discussed in the previous paragraph about the representativeness 
of the solar $^{12}$C/$^{13}$C ratio also apply to the $^{14}$N/$^{15}$N ratio. 
Accurate measurements of the $^{14}$N/$^{15}$N ratios toward warm molecular 
clouds spanning a range of Galactocentric distances have been obtained by 
\cite{2012ApJ...744..194A} from millimetre-wave observations of rotational 
lines of CN and HNC and their isotopologues. Direct (from CN, correcting for 
opacities when needed) and indirect \citep[from HNC, using the $^{12}$C/$^{13}$C 
ratios previously established by][for each source]{2005ApJ...634.1126M} 
determinations yield the same gradient, within the uncertainties. The 
$^{14}$N/$^{15}$N ratios derived by \cite{2012ApJ...744..194A}, however, are 
systematically lower than those obtained by \cite{1995A&A...295..194D} from 
H$^{13}$CN/HC$^{15}$N data, likely because of the use of $^{12}$C/$^{13}$C ratios 
from H$_2$CO in \cite{1995A&A...295..194D}.  Indeed, scaling the HCN data of 
\cite{1995A&A...295..194D} with $^{12}$C/$^{13}$C ratios from CN yields HCN 
values that agree with those from the other indicators, within the 
uncertainties \citep[see figure~3 of][]{2012ApJ...744..194A}. It is worth 
noting that, while \cite{2003MNRAS.342..185R} had to offset their model 
predictions to match \cite{1995A&A...295..194D} data, the new estimates of the 
$^{14}$N/$^{15}$N ratio across the Galaxy by \cite{2012ApJ...744..194A} make 
such a correction unnecessary. The mean local ISM value suggested by 
\cite{2012ApJ...744..194A}, ${\rm ^{14}N/^{15}N} =290\pm 40$, agrees with that 
measured in nearby diffuse clouds from CN absorption lines in the optical 
\citep[274$\pm$18;][]{2015ApJ...804L...3R}.

We adopt the solar photospheric ratios ${\rm ^{16}O/^{18}O} = 511 \pm 10$ and 
${\rm ^{18}O/^{17}O} =5.36 \pm 0.34$ from \cite{2013ApJ...765...46A} as 
indicative of the local ISM composition 4.5\,Gyr ago. A radial $^{18}$O/$^{17}$O 
gradient is suggested by \cite{2008A&A...487..237W}, who combine observations 
of different CO transitions in Galactic sources covering the Galactocentric 
distance range $0< R_{\mathrm{GC}}/{\rm kpc}< 17$. Their finding is further 
supported by recent independent analysis and observations by 
\cite{2016RAA....16...47L}. Ratios of $^{16}$O to $^{18}$O across the disc are 
taken from the compilation of \cite{1994ARA&A..32..191W}, from OH data of 
\cite{2005A&A...437..957P}, and derived by combining the CO data from 
\cite{2008A&A...487..237W} with the $^{12}$C/$^{13}$C gradient from 
\cite{2005ApJ...634.1126M} following \cite{2011ApJ...729...43Y}.

\subsection{Other galaxies}
\label{sec:EXTdata}

CNO isotopic abundances can be precisely measured on the Earth and in the Sun. 
However, they are more difficult to obtain in the Galactic ISM and --even more 
so-- in the extragalactic ISM, due to the limits of past observational 
capabilities and the high optical depths of the molecular lines of the most 
abundant isotopologues. Because of past sensitivity limitations, measurements 
of the extragalactic isotopic CO transitions have been limited to local 
gas-rich and metal-enriched galaxies \citep[e.g.][]{2001ApJS..135..183P,
2011A&A...528A..30C, 2011RAA....11..787T, 2014MNRAS.445.2378D}. With the help 
of gravitational lensing, which can greatly amplify the flux densities, 
isotopic lines at $z > 2$ have now been measured in a few cases 
\citep{2006A&A...458..417M, 2010A&A...516A.111H, 2013MNRAS.436.2793D}.

Isotopic abundances are difficult to derive from the measured molecular lines, 
especially for the $^{12}$C-bearing species, which are the most abundant and 
thus their  transitions have the highest opacities. Observed line ratios do not 
translate directly to the abundance ratios --one needs to correct for the 
unknown optical depths. As a result, the $^{12}$C/$^{13}$C and $^{16}$O/$^{18}$O 
abundance ratios from the line ratios are often significantly underestimated. 
The most straightforward method to derive the optical depths and obtain the 
isotopic abundance ratios uses the line ratios between the optically thin 
double isotopologue (e.g.\ $^{13}$C$^{18}$O) and a single isotopic molecular 
line (e.g.  C$^{18}$O) of the same quantum transition. However, should the 
isotopologue line (e.g., C$^{18}$O) not be completely optically thin, would 
mean an underestimate of the derived $^{12}$C/$^{13}$C abundance ratio. 
Moreover, the emission lines from the double isotopologues are so weak that it 
has only been possible to detect them in a few bright targets, even with the 
most sensitive radio telescopes \citep[e.g.\ M\,82, 
NGC\,253;][]{2010A&A...522A..62M}. ALMA will improve this situation 
considerably, but a large survey of nearby normal galaxies, or high-redshift 
star-forming galaxies, will remain difficult, unless lensed objects are used in 
the later case.

A second method derives the abundance ratios using the absorption features of 
molecular isotopologues against strong radio continuum sources. This is perhaps 
the most accurate method to derive the column densities from the measured 
equivalent widths, as measured directly from the isotopologues. This method is 
also insensitive to distance and angular resolution. Its main drawback, 
however, is that a very strong background source is needed.  Only the 
combination of galactic ISM in the line of sight to the brightest radio-loud 
quasars, or very strong continuum emission from the target itself, is adequate 
for such studies. This method has been used to measure the diffuse ISM in our 
Milky Way galaxy \citep[e.g.][]{1996A&A...307..237L, 1998A&A...337..246L}, 
local galaxies \citep[e.g.][]{2012A&A...541A...4G} and a few distant galaxies 
\citep[e.g.][]{1995A&A...299..382W, 1996A&A...315...86W, 2014A&A...566A.112M}. 
Targets lending themselves to such studies are thin on the ground because of 
the rare configuration requirements. Optical observations of 
Damped Lyman-$\alpha$ system (DLA) absorbers also present the ability to probe 
isotopologue ratios in high-$z$ objects \citep{2006A&A...447L..21L, 
2017A&A...597A..82N}.

A third method requires two steps, deriving the optical depths and the 
abundance ratios separately using the N-bearing molecular line. Most often, 
$^{13}$CN, $^{12}$CN $N=1-0$ and its hyperfine structure lines are observed, and 
the optical depth of $^{12}$CN is derived. Then the opacity-corrected $^{12}$CN 
is compared with the emission of $^{13}$CN, which is assumed to be optically 
thin. Their ratio yields the $^{12}$C/$^{13}$C abundance ratio, which can be 
further converted to the $^{16}$O/$^{18}$O ratio using observations of $^{13}$CO 
and C$^{18}$O \citep[e.g.][]{2002ApJ...578..211S,2005ApJ...634.1126M}. However, 
this method relies on measurements of the CN molecule, which requires 
high-density conditions \citep[$n_{\rm crit} ^{\rm CN\,1-0} \sim 
10^{4-5}$\,cm$^{-3}$;][]{2015PASP..127..299S} like those found in the dense cores 
of star-forming regions, where optical depths are also high. Furthermore, the 
assumption of an optically thin $^{13}$CO may not always hold, as with the 
first method.

A fourth method models the average gas physical conditions using multiple 
rotational transitions of the isotopologues and radiative transfer models based 
on the large velocity gradient (LVG) approximation \citep{1974ApJ...187L..67S,
1974ApJ...189..441G} or the mean-escape probability (MEP) approximation 
\citep{1989agna.book.....O, 2001A&A...367..457C}.  This allows an estimate of 
the optical depths of both the major and minor isotopologues, which allows the 
appropriate correction of the observed line ratios to get the abundance ratios. 
It is much easier to obtain multiple transitions of CO isotopologues rather 
than for other even rarer molecules. However, because of the degeneracy of the 
collisional coefficient between density and temperature, $C \propto n_{\rm H_2} 
T_{\rm kin}^{1/2}$, the degeneracy between molecular abundances and other gas 
properties (e.g. average $dV/dR$) that set a given line optical depth, and the 
assumption of uniformly distributed physical conditions in studied regions, the 
uncertainty of the deduced abundance ratio can be considerable. Moreover, in 
the case of the CO/$^{13}$CO (or CO/C$^{18}$O)  line ratio, the stronger 
radiative trapping expected for the much more abundant CO than for $^{13}$CO 
can also play some role in boosting their values in some galaxies in addition 
to enhanced global CO/$^{13}$CO abundances in their ISM 
\citep[e.g.,][]{1995A&A...300..369A}.

In Table~\ref{tab:ext} we list the abundance ratios published in the 
literature. There are many detections of a single transition of $^{13}$CO in 
galaxies, both in their centres and in the off-nuclear arm/disc regions 
\citep[e.g.][]{2001ApJS..135..183P, 2011RAA....11..787T, 2014MNRAS.445.2378D}. 
In this study, we adopt only the abundance ratios; we neglect the line ratios 
which, in principle, could be used to set the lower limits.

\subsection{From isotopologue to isotope abundance ratios: the road is now 
  clear}
\label{sec:fromto}

The first detections of isotopologue $^{13}$CO and C$^{18}$O line emission in 
the Galaxy \citep{1971ApJ...165..229P} were soon followed by numerous 
detections of other, more distant galaxies \citep[][]{1979A&A....78L...1E,
1985ApJ...292L..57R,1986ApJ...302..680Y,1991A&A...247..320S,1993ApJ...414...98W,
1995A&A...300..369A,1996ApJ...465..173P, 1998ApJ...492..521P,
2001ApJS..135..183P}. This facilitated the use of isotopologue abundance ratios 
as direct measures of isotope abundance ratios in the ISM of other galaxies 
\citep[see][for Galactic studies of the $^{13}$CO, $^{12}$CO 
isotopologues]{1993ApJ...408..539L}. It was this new observational capability, 
and the discovery that $^{12}$CO/$^{13}$CO {\it intensity} ratios are typically 
much larger in extreme merger/starburst galaxies than in ordinary star-forming 
spirals \citep[][]{1992A&A...264...55C, 1993A&A...274..730H,
1993A&A...268L..17H, 1995A&A...300..369A,1998ApJ...492..521P} which triggered 
investigations into whether or not isotopologue abundance ratios translate 
directly to isotopic ratios (e.g.\ [$^{12}$CO/$^{13}$CO] = [$^{12}$C/$^{13}$C]).

Two astrochemical effects stand in the way of obtaining an isotope abundance 
ratio from the corresponding isotopologue abundance ratio, namely: 1) selective 
photodissociation of the rarer isotope \citep{1992A&A...264...55C} and 2) 
isotope chemical fractionation \citep{1984ApJ...277..581L,2013A&A...550A..56R}. 
The first operates in the far-ultraviolet (FUV)-illuminated, warm outer layers 
of molecular clouds (the so-called `photodissociation regions', PDRs); the 
second operates in the cold, FUV-shielded inner regions of those molecular 
clouds. In past work selective photodissociation of $^{13}$CO with respect to 
$^{12}$CO has been investigated as the cause of the high $^{12}$CO/$^{13}$CO 
intensity ratios in (ultra)luminous infrared galaxies [(U)LIRGs], but found 
unlikely for large masses of their molecular gas reservoirs 
\citep{1992A&A...264...55C}.

This can be shown for the metal-rich, high-pressure molecular gas reservoirs 
expected in merger/starbursts such local (U)LIRGs and distant submm-selected 
galaxies (SMGs). Indeed in such metal-rich environments it can be shown that 
the $\rm H_2$ gas mass fraction expected to be in PDRs (the only gas that 
could be affected by selective photodissociation effects) is small, and thus 
can not perturb global isotopologue abundance ratios away from the elemental 
isotopic abundance ratios. Indeed, following  the method developed in 
\citet[][section~5.1, Equation~6]{2014ApJ...788..153P},

\begin{equation}
        f_{\rm PDR} \sim 2 \times [ 1 - ( 1- \frac {4 A_v^{\rm tr}}{3 <A_v>})^3 ],
\end{equation}

\noindent where $A_v^{\rm tr} = 1.086\, \xi_{\rm FUV}^{-1} {\rm ln }[ 1 + \Phi\, 
G_0^{\rm FUV}\, k_0 / (n R_{\rm f})] $ is the visual extinction corresponding to a 
transition (between H{\sc i} and H$_2$) column density, $\xi_{\rm FUV} = 
\sigma_{\rm FUV} / \sigma_{\rm V} \sim 2-3$ is the dust cross section ratio for 
FUV and optical lights, $\Phi = 6.6 \times 10^{-6}\sqrt{\pi} Z^{1/2} \xi_{\rm FUV}$ 
is the H$_2$ self-shielding fraction over the H{\sc i}/H$_2$ transition layer, 
$G_0^{\rm FUV}$ is the PDR radiation field, $k_0 = 4 \times 10^{-11}$s$^{-1}$ is 
the H$_2$ dissociation rate, $n$ is the average gas density, $R_{\rm f} \sim 3 
\times 10^{-17}\rm cm^{-3}$ s$^{-1}$, and $Z=1$ (solar) is the metallicity.

We find the PDR gas mass fraction per cloud to be $f_{\rm PDR}\sim 0.004$--0.06 
(H\,{\sc i}+$\rm H_2$) for the metal-rich and high-pressure (and thus 
high-density) molecular gas of (U)LIRGs. Even for less extreme conditions, with 
$n \sim 5\times 10^3$\,cm$^{-3}$, $T_{\rm kin} \sim 20$\,{\sc K}, 
$P_{\rm e}/k_{\rm B} \sim 10^5$\,cm$^{-3}$, we obtain $f_{\rm PDR} \sim 0.11$--0.17, 
of which only around half will be molecular gas, and a smaller fraction still 
will be affected by selective photodissociation of the rarer isotopologues. 
These small gas mass fractions ensure that {\it selective photodissociation 
cannot play a major role for the bulk of the  molecular gas found in 
dust-enshrouded starbursts.} Finally, the fact that cloud-volumetric heating 
processes such as cosmic rays (CRs) and/or X-rays rather than (cloud 
surface)-heating FUV photons of PDRs seem to be responsible for the average 
thermal state of molecular gas in LIRGs 
\citep[e.g.,][]{2010A&A...518L..37P,2010A&A...518L..42V,
2012MNRAS.426.2601P,2017ApJ...836..117I}, makes it even more unlikely that a 
FUV-driven isotopologue selective dissociation process affects much of their 
molecular gas mass reservoirs.

At the cold end, isotope fractionation operates via exothermic isotope-exchange 
chemical reactions whose $T_{\rm kin}$-sensitivity lies in a range where 
molecular gas is found in ordinary spirals. Indeed, the three most important 
reactions behind the theory of $^{12}$C/$^{13}$C fractionation are:
\begin{equation}
 ^{13}{\rm C}^+ + {\rm CO} \rightleftharpoons {\rm C}^+ + {\rm ^{13}CO} + 
  35\,{\rm K,}
\end{equation}
\begin{equation}
 ^{13}{\rm CO} + {\rm HCO}^+ \rightleftharpoons {\rm CO} + 
  {\rm H}^{13}{\rm CO}^+ + 17\,{\rm K,}
\end{equation}
\begin{equation}
 ^{13}{\rm C}^+ + {\rm CN} \rightleftharpoons {\rm C}^+ + {\rm ^{13}CN} + 
  31\,{\rm K}
\end{equation}
\noindent \citep[][and references therein]{2013A&A...550A..56R,
2015ApJ...815..114T}. The effects of isotope fractionation have recently been 
invoked to explain the isotopologue $^{13}$CO, $^{12}$CO and H$^{13}$CN, 
H$^{12}$CN line ratios observed in the (U)LIRG, NGC\,6240 
\citep{2015ApJ...815..114T}. However, this work did not consider the minimum 
$T_{\rm kin}$ set by the CR heating that is expected to be ubiquitous in the 
environments of starbursts, and argued that dense ($n=10^6$\,cm$^{-3}$) and very 
cold ($T_{\rm kin}= 10$\,{\sc K}) gas comprises the most massive phase of 
$\rm H_2$ in NGC\,6240. At such low temperatures, isotopic fractionation can 
indeed operate \citep[e.g.][]{2013A&A...550A..56R}, but these conditions can 
only be found in dense gas cores, deep inside giant molecular clouds in the 
relatively quiescent Milky Way, with its low levels of average CR energy 
density.

Using figure~1 from \cite{2011MNRAS.414.1705P}, we find $T_{\rm kin} 
=20$--30\,{\sc K} is the minimum value set by CRs in FUV-shielded dense gas in 
(U)LIRGs, where CR energy densities are 100--500$\times$ that in the Galaxy and 
$n= 10^5$\,cm$^{-3}$. This temperature can be higher still if turbulent gas 
heating and/or heating from IR-heated dust remains significant in such regions. 
Moreover, the discovery of high HCN($J=4-3$) brightness temperatures in 
high-resolution ALMA maps of the (U)LIRG, Arp\,220 (Project: 2015.1.00702S, PI: 
L.~Barcos-Munoz), with $T_{\rm b}\,{\rm HCN}\,(J=1-0)\sim 70$--80\,{\sc K} sets 
a lower limit for $T_{\rm kin}$ in the dense, line-emitting HCN gas that is well 
above the regime where isotope fractionation can operate.

For NGC\,6240, ALMA observations (Project: 2012.00077.S, PI: N.~Scoville) give 
$T_{\rm b}\,{\rm  HCN}\,(J=4-3)\sim 3$\,{\sc  K} at 70-pc spatial resolution; 
thus for its observed HCN($4-3/1-0$) global brightness ratio of 0.18 it would 
be: $T_{\rm b}\,{\rm  HCN}\,(J=1-0)\sim 17$\,{\sc K} as the corresponding lower 
limit for the temperature of the dense gas. This is well above the temperature 
range where chemical fractionation can operate. On the other hand, the 
$T_{\rm dust}\sim  40$\,{\sc K} found in NGC\,6240 \citep{2000MNRAS.312..433L}, 
and for a normal gas-to-dust mass ratio, there is no room for a significant 
mass of colder dust. Since $T_{\rm kin} \ge T_{\rm dust}$ in the bulk of the ISM 
(irrespective of whether the energetics are FUV- and/or CR-driven), a scenario 
involving large amounts of gas at low temperatures and high densities (where 
chemical fractionation can operate) is highly unlikely. \emph{We thus conclude 
that isotope chemical fractionation cannot operate for the bulk of the 
molecular gas in star-forming galaxies.}

Thus the road is now clear to use global isotopologue abundance ratios to 
obtain the corresponding average isotope abundance ratios in the ISM of 
galaxies. Nevertheless, serious uncertainties remain, associated with the 
radiative transfer modeling of molecular line emission. Indeed, careful 
radiative transfer modeling is needed to obtain the range of isotopologue 
abundance ratios, such as [$^{12}$CO/$^{13}$CO], from multi-$J$ $^{12}$CO and 
$^{13}$CO emission line observations. Even when multiple lines are available, 
degeneracies still remain \citep[e.g.][]{2014ApJ...788..153P}. Better angular 
resolution is also needed to resolve the local variation of the isotopologue 
line ratios, which can be resolved easily with ALMA now 
\citep[e.g.,][]{2017ApJ...836L..29J}. In the age of ALMA, with its 
extraordinary sensitivity, angular resolution and flexible, wide-band 
correlator, simultaneous isotopologue line observations will allow such 
degeneracies to be considerably reduced. Moreover, using multi-$J$ \emph{and} 
multi-species isotopologue line observations to determine a single isotope 
ratio \citep[e.g.~$^{12}$CO, $^{13}$CO and H$^{12}$CN, H$^{13}$CN lines to 
determine $^{12}$C/$^{13}$C --][]{1997ApJ...484..656P} will further reduce the 
model degeneracies.

\section{Chemical evolution models}
\label{sec:mod}

%%%%%%%%%%%%%%%%%%%%%%%%%%%%%%%%%%%%%%%%%%%%%%%% 
%TWO COLUMN TABLE

\begin{table*}
\caption{Prescriptions for nucleosynthesis.}
\begin{tabular}{@{}lcccc@{}}
\hline
Model & LIMS & Super-AGB stars & Massive stars & Novae \\
\hline
1 & Karakas (2010) & -- & Nomoto et al.~(2013) & No \\
2 & Karakas (2010) & Doherty et al.~(2014a,b) & Nomoto et al.~(2013) & No \\
3 & Karakas (2010) & -- & Meynet \& Maeder (2002b), Hirschi et al.~(2005), Hirschi (2007), Ekstr\"om et al.~(2008) & No \\
4 & Karakas (2010) & Doherty et al.~(2014a,b) & Meynet \& Maeder (2002b), Hirschi et al.~(2005), Hirschi (2007), Ekstr\"om et al.~(2008) & No \\
5 & Karakas (2010) & Doherty et al.~(2014a,b) & Nomoto et al.~(2013) & Yes \\
\hline
\end{tabular}
\label{tab:nuc}
%       \begin{flushleft}
%       \emph{Notes.}
%       \end{flushleft}
\end{table*}

%%%%%%%%%%%%%%%%%%%%%%%%%%%%%%%%%%%%%%%%%%%%%%%%

The chemical evolution model for the Milky Way used in this study was 
established in a series of papers (\citealt{1997ApJ...477..765C,
2001ApJ...554.1044C, 2000ApJ...539..235R, 2010A&A...522A..32R,
2015ApJ...802..129S}; see also \citealt{1986A&A...154..279M,
1989MNRAS.239..885M}) to which we refer the reader for a thorough discussion of 
the adopted formalism, the basic equations and the assumptions. It is a 
multi-zone model, where the Galactic disc is divided in several concentric 
annuli that evolve at different rates; this ensures the establishment of a 
Galactic abundance gradient.

For the other galaxies, we run simpler single-zone models. We do not aim to 
reproduce the CNO isotopic data available for each specific object; rather, we 
seek to give a broad overview of how the CNO isotopic ratios are expected to 
evolve in systems with different star-formation histories and/or stellar IMFs. 
In these models, fresh gas is accreted according to an exponentially-decreasing 
law, d$\mathscr{M}_{\rm{inf}}$/d$t \propto \rm{e}^{-t/\tau}$ --where 
$\mathscr{M}_{\rm{inf}}$ is the total mass accreted and $\tau$ is the infall 
timescale-- and turned into stars following a Kennicutt-Schmidt relation, 
$\psi(t) = \nu\,\mathscr{M}_{\rm{gas}}(t)$, where $\nu$ is the star-formation 
efficiency and $\mathscr{M}_{\rm{gas}}$ is the mass of neutral gas
\citep{1959ApJ...129..243S, 1998ApJ...498..541K}. The free parameters, $\tau$ 
and $\nu$, are set to different values in order to frame different evolutionary 
paths. A more detailed description of the adopted formalism can be found in 
section~3.1 of \cite{2015MNRAS.446.4220R}.

Specifically, we consider: (i) a template for massive systems 
($\mathscr{M}_{\mathrm{DM}} =10^{13}$\,M$_\odot$, $\mathscr{M}_\star \sim 2\times 
10^{11}$\,M$_\odot$) that accrete gas rapidly ($\tau =0.05$\,Gyr) and experience 
powerful starbursts ($\nu \simeq1$--2\,Gyr$^{-1}$) at high redshift, followed 
by passive evolution thereafter\footnote{These objects would appear as sub-mm 
  galaxies \citep[SMGs; e.g.][]{1997ApJ...490L...5S} at redshifts 
  $z \simeq 2$--3 and as `red and dead' massive ellipticals at $z=0$ 
  \citep[see, e.g.,][]{2014ApJ...782...68T}.}; (ii) a template for massive 
spirals ($\mathscr{M}_{\mathrm{DM}} \simeq 10^{12}$\,M$_\odot$, $\mathscr{M}_\star 
\sim 5\times 10^{10}$\,M$_\odot$) that accrete gas slowly ($\tau =13$\,Gyr) and 
experience steady star formation ($\nu \simeq 0.05$--0.1\,Gyr$^{-1}$) over a 
Hubble time; (iii) a model similar to the latter, but for galaxies of lower 
mass ($\mathscr{M}_{\mathrm{DM}} \simeq 10^{11}$\,M$_\odot$, $\mathscr{M}_\star \sim 
10^{10}$\,M$_\odot$) where secular evolution is followed suddenly by an extremely 
efficient ($\nu \simeq 20$\,Gyr$^{-1}$) burst of star formation. The 
star-formation histories of these systems are shown in the top panels of 
Fig.~\ref{fig:ext}, for different choices of the stellar IMF (see next 
paragraph; see also discussion in \S\ref{sec:EXTres}).

%%%%%%%%%%%%%%%%%%%%%%%%%%%%%%%%%%%%%%%%%%%%%%%%
%TWO COLUMN FIGURE

\begin{figure*}
\begin{center}
\includegraphics[width=\columnwidth]{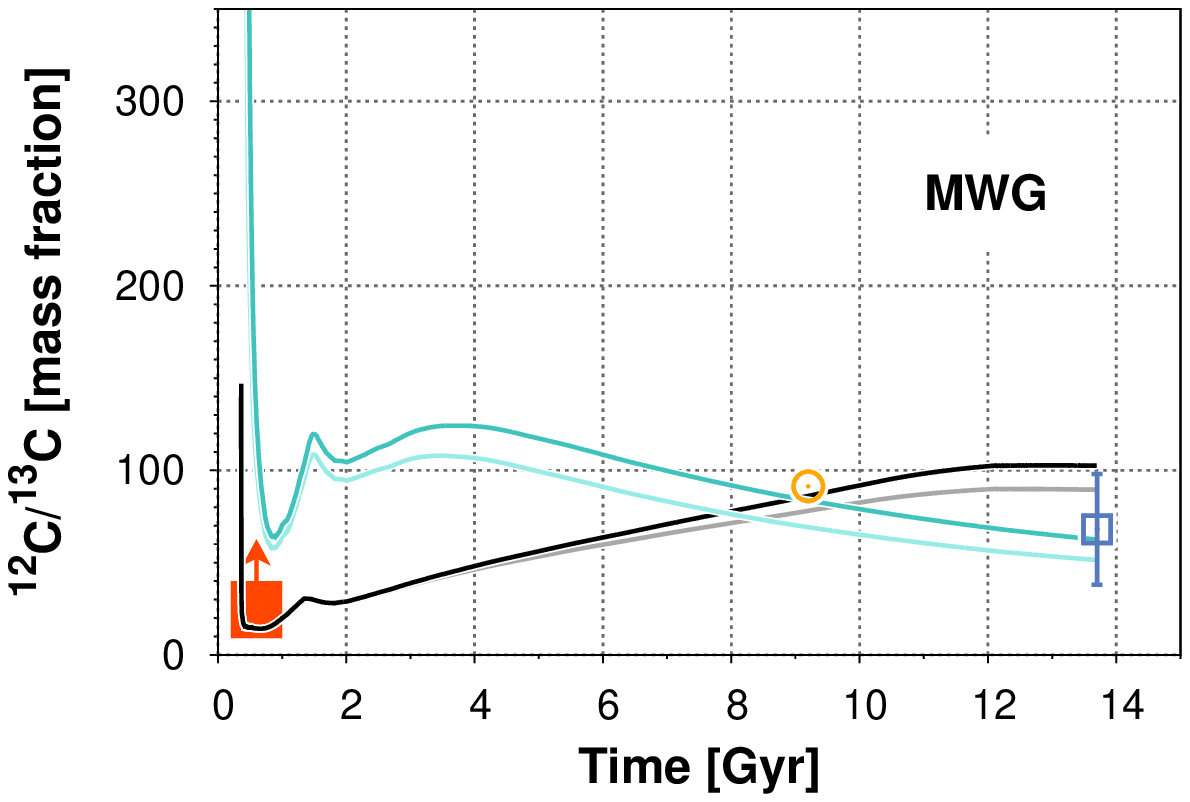} 
\includegraphics[width=\columnwidth]{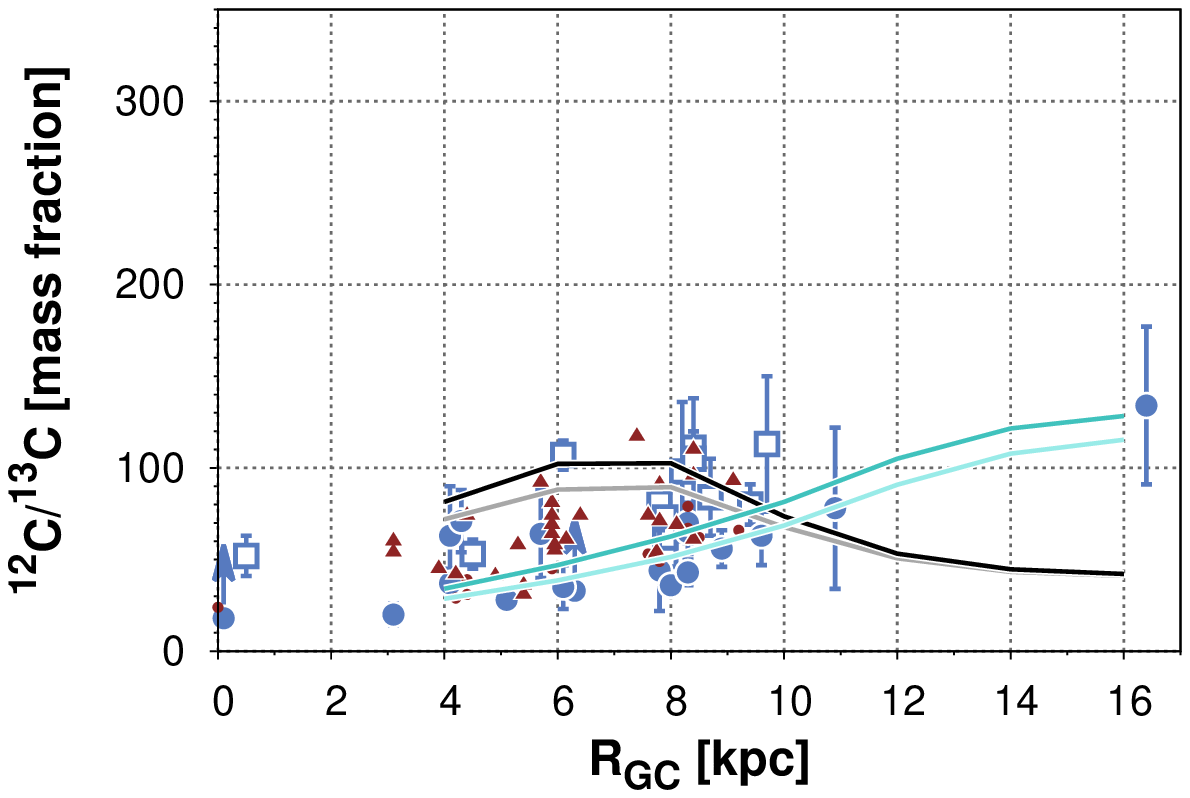}
\caption{ \emph{Left panel:} evolution of the carbon isotope ratio in the solar 
  neighbourhood, predicted by Models~1 (green line), 2 (light green line), 3 
  (black line) and 4 (grey line). The red box encompasses the $^{12}$C/$^{13}$C 
  ratios of `unmixed' halo stars \citep{2006A&A...455..291S}. The solar 
  \citep[$91.4\pm 1.3$;][]{2013ApJ...765...46A} and local ISM values 
  \citep[$68\pm 15$;][]{2005ApJ...634.1126M} are denoted by different 
  symbols and colours. \emph{Right panel:} radial behaviour of the 
  $^{12}$C/$^{13}$C ratio at the present time predicted by Models 1 (green 
  line), 2 (light green line), 3 (black line) and 4 (grey line). 
  $^{12}$C/$^{13}$C ratios across the disc inferred from observations of CO, 
  H$_2$CO \citep[as taken from the compilation of][]{1994ARA&A..32..191W}, 
  CO$_2$ \citep{2000A&A...353..349B} and CN \citep{2002ApJ...578..211S,
    2005ApJ...634.1126M} are shown as dots, filled triangles, open squares and 
  filled circles, respectively.}
\label{fig:12c13cMWG}
\end{center}
\end{figure*}

%%%%%%%%%%%%%%%%%%%%%%%%%%%%%%%%%%%%%%%%%%%%%%%%

The probability that a newly born star has an initial mass within a given mass 
range is given by the \cite{2002ASPC..285...86K} IMF, with a slope $x =1.7$ for 
high masses, normalised to unity in the 0.1--100-M$_\odot$ range. For the 
starbursts, we also investigate the effects of an IMF skewed toward high masses 
($x = 0.95$ in the high-mass regime), similar to the one proposed by 
\cite{2007A&A...467..123B} for the Galactic bulge.

All our computations avoid the instantaneous recycling approximation by 
detailed accounting of the finite stellar lifetimes. This is a necessary 
prerequisite for a proper treatment of elements that are produced on different 
timescales by stars of different initial masses and chemical compositions. The 
nucleosynthetic outcomes of binary stars exploding as type Ia supernovae (SNe 
Ia) and novae are included in our computations, adopting the single-degenerate 
scenario for their progenitors \citep[][and references 
therein]{2001ApJ...558..351M} for SNe Ia and our previous work 
(\citealt{1999A&A...352..117R, 2001A&A...374..646R}; based on 
\citealt{1991A&A...248...62D}) for novae. Since CNO elements are produced in 
negligible amounts in SN Ia explosions \citep{1999ApJS..125..439I}, the exact 
choice of the route leading to such events does not affect the results 
presented in this paper. Classical novae, instead, are thought to significantly 
overproduce $^{13}$C, $^{15}$N and $^{17}$O with respect to their solar 
abundances \citep[e.g.][]{2007JPhG...34..431J} making assumptions about their 
precursors a much more thorny problem (see next section).

%%%%%%%%%%%%%%%%%%%%%%%%%%%%%%%%%%%%%%%%%%%%%%%%
%TWO COLUMN FIGURE

\begin{figure*}
\begin{center}
\includegraphics[width=\columnwidth]{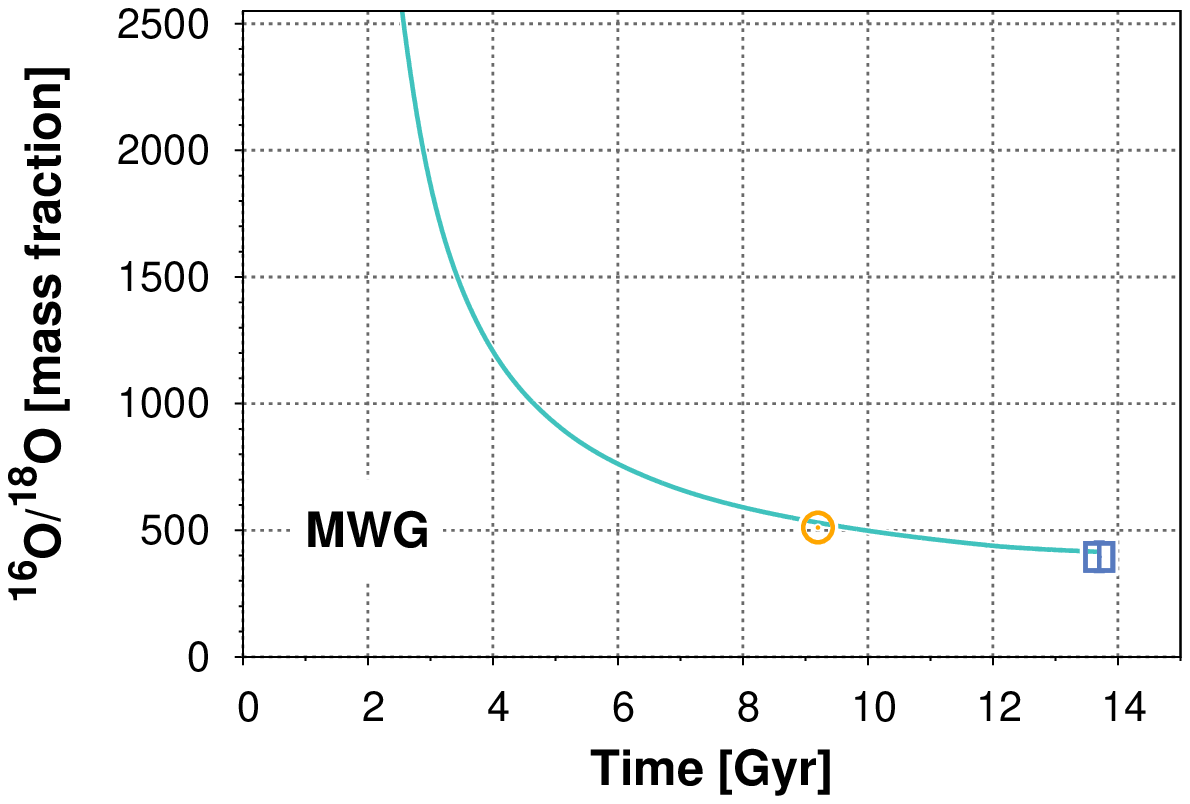}
\includegraphics[width=\columnwidth]{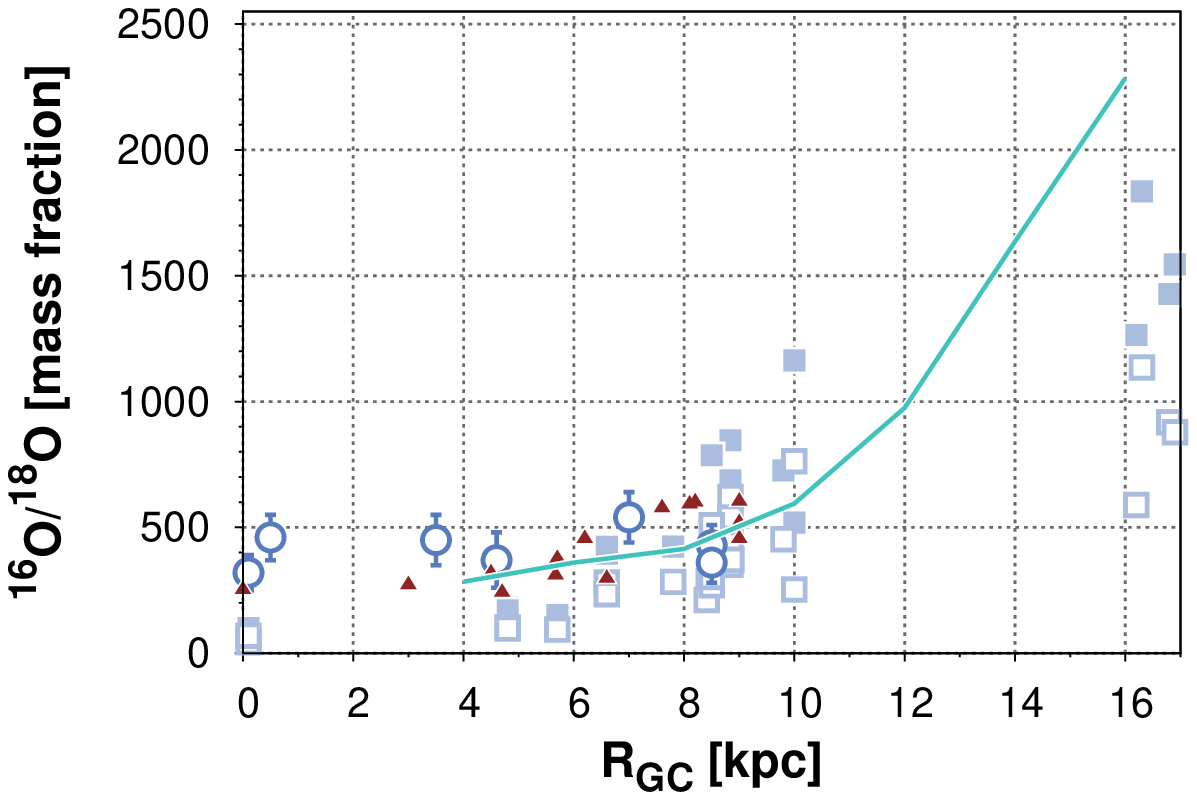}
\caption{{\it Left panel:} evolution of $^{16}$O/$^{18}$O in the solar 
  neighbourhood according to the predictions of Models~1 (green solid line) and 
  2 (light green line, hidden behind Model~1 predictions). Solar 
  \citep[$511\pm 10$;][]{2013ApJ...765...46A} and local ISM values 
  \citep[$395\pm 56$;][]{2005A&A...437..957P} are shown. \emph{Right panel:} 
  current ratios of $^{16}$O to $^{18}$O against distance from the Galactic 
  centre predicted by Models~1 (green solid line) and 2 (light green line, 
  hidden behind Model~1 predictions). Theoretical expectations are contrasted 
  with formaldehyde data from the compilation of 
  \citet[][triangles]{1994ARA&A..32..191W}, OH data from \citet[][open 
  circles]{2005A&A...437..957P} and CO data obtained by combining 
  $^{13}$C$^{16}$O/$^{12}$C$^{18}$O line ratios from \citet{2008A&A...487..237W} 
  with the $^{12}$C/$^{13}$C gradient suggested by \citet[][filled squares: 
    $J=1-0$ transition; open squares: $J=2-1$ transition]{2005ApJ...634.1126M}.}
\label{fig:16o18oMWG}
\end{center}
\end{figure*}

%%%%%%%%%%%%%%%%%%%%%%%%%%%%%%%%%%%%%%%%%%%%%%%%
   
%%%%%%%%%%%%%%%%%%%%%%%%%%%%%%%%%%%%%%%%%%%%%%%%
%TWO COLUMN FIGURE

\begin{figure*}
\begin{center}
\includegraphics[width=\columnwidth]{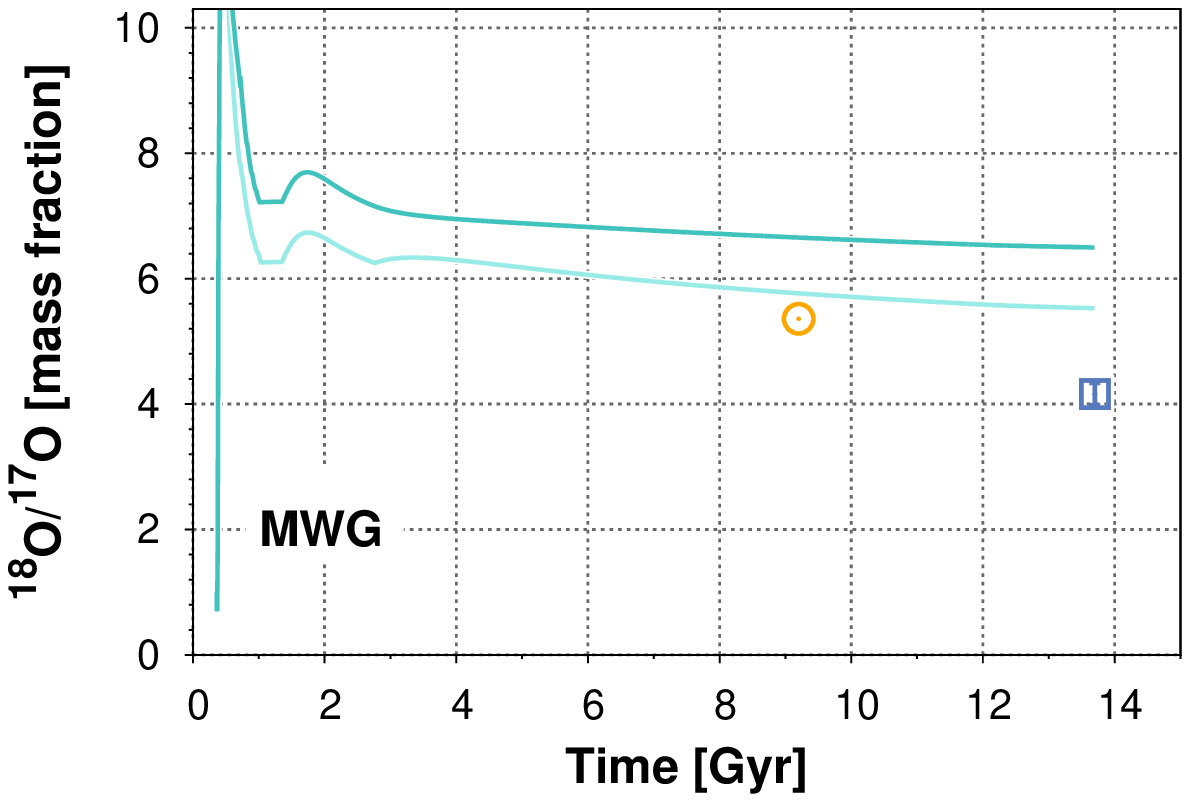}
\includegraphics[width=\columnwidth]{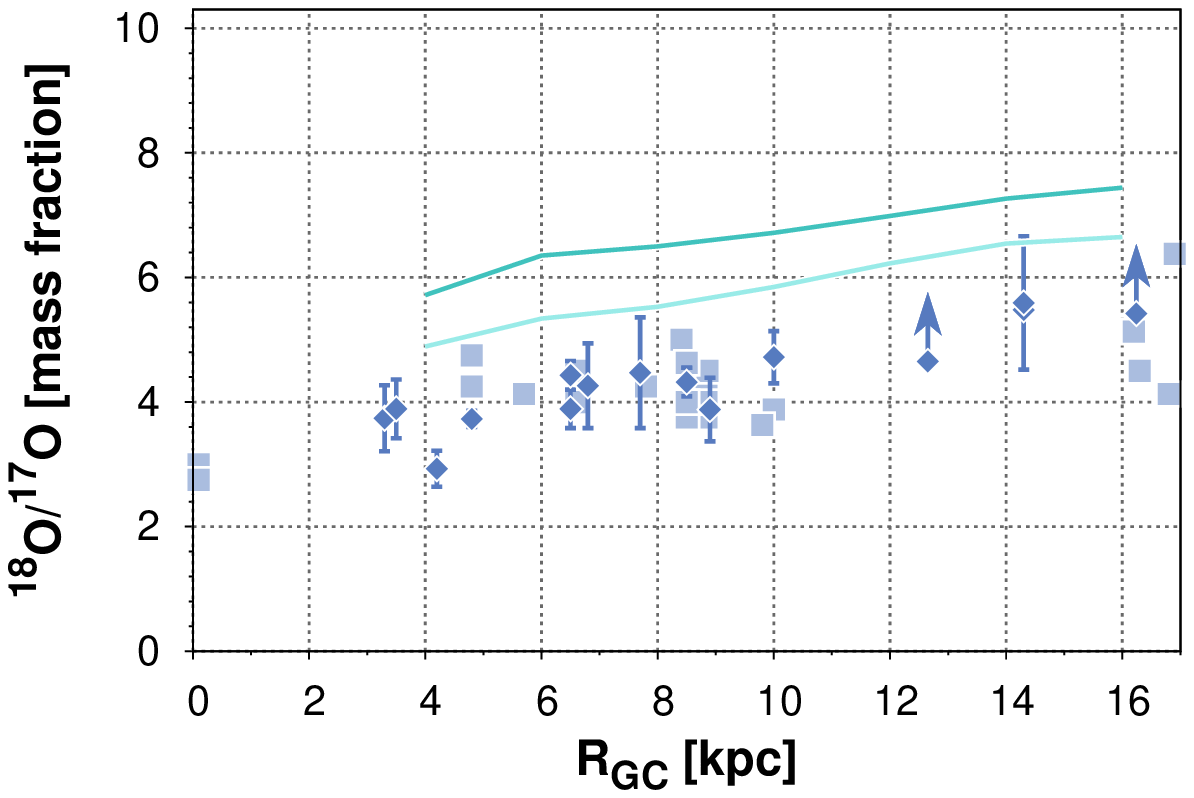}
\caption{{\it Left panel:} evolution of $^{18}$O/$^{17}$O in the solar 
  neighbourhood predicted by Models~1 (green line) and 2 (light green line) 
  compared to the solar \citep[$5.36\pm 0.34$, Sun 
    symbol;][]{2013ApJ...765...46A} and local ISM values \citep[$4.16\pm 0.09$, 
    open square;][]{2008A&A...487..237W}. \emph{Right panel:} $^{18}$O/$^{17}$O 
  ratios across the disc at the present time. Theoretical predictions are shown 
  together with ratios derived from CO lines by \citet[][filled 
    squares]{2008A&A...487..237W} and \citet[][diamonds]{2016RAA....16...47L}.}
\label{fig:18o17oMWG}
\end{center}
\end{figure*}

%%%%%%%%%%%%%%%%%%%%%%%%%%%%%%%%%%%%%%%%%%%%%%%%

The adopted nucleosynthesis prescriptions are summarised in 
Table~\ref{tab:nuc}. The stellar yields for low- and intermediate-mass stars 
are from \cite{2010MNRAS.403.1413K}. For massive stars, we adopt the grid of 
yields suggested by \citet[][Models~1, 2 and 5]{2013ARA&A..51..457N}. However, 
since the important effects of stellar rotation are not accounted for in this 
case, we also consider the grid of CNO yields from fast-rotating massive stars 
provided by the Geneva group \citep[][Models~3 and 4]{2002A&A...390..561M,
2005A&A...433.1013H, 2007A&A...461..571H,2008A&A...489..685E}. Detailed yields 
for super-AGB stars \citep{2014MNRAS.437..195D,2014MNRAS.441..582D} --often 
neglected in chemical evolution studies-- are implemented in Models~2, 4 and 5,
while nova nucleosynthesis --also disregarded in most studies-- is included in 
one case (Model~5). All the adopted yields for single stars are dependent on 
mass and metallicity. For novae, we assume average yields, independent of mass 
and metallicity (see discussion in \S\ref{sec:MWres}). The different sets of 
yields were all tested against the Milky Way data, while only the set labeled 
`1' was used in the models for the other galaxies.

\section{Results}
\label{sec:res}

\subsection{Milky Way galaxy}
\label{sec:MWres}

In this section, we present the results of our chemical evolution models for 
the Milky Way. These models differ in their adopted nucleosynthesis 
prescriptions (see Table~\ref{tab:nuc}). First, we discuss the predictions of 
Models~1, 2, 3 and 4, where the CNO elements come only from single stars. We 
then analyse the results of Model~5, which includes CNO production during nova 
outbursts.

Fig.~\ref{fig:12c13cMWG} shows the evolution of the carbon isotope ratio in the 
solar neighbourhood (left panel) and its current behaviour across the Milky Way 
disc (right panel) predicted by Models~1 (green lines), 2 (light green lines), 
3 (black lines) and 4 (grey lines).  Models~1 and 3, which do not include the 
contribution to C synthesis from super-AGB stars, successfully reproduce the 
solar data \citep{2013ApJ...765...46A}; adding the super-AGB star contribution, 
the solar $^{12}$C/$^{13}$C ratio is slightly underestimated (Models~2 and 4). 
However, one must be aware that the Sun probably moved to its current position 
from a birthplace closer to the Galactic centre \citep{1996A&A...314..438W} and 
its chemical composition may not quite reflect that of the local ISM 4.5\,Gyr 
ago. All the models predict current $^{12}$C/$^{13}$C ratios in the solar 
neighbourhood that agree with the average local ISM value 
\citep{2005ApJ...634.1126M}, within the errors.  Furthermore, Models~3 and 4, 
assuming CNO yields from fast-rotating massive stars, can account for the range 
of carbon isotope ratios of `unmixed' halo stars \citep[i.e.\ giant stars in 
which the mixing with the deep layers affecting the original CNO abundances is 
not expected to have occurred;][]{2006A&A...455..291S}. This result was already 
discussed by \cite{2008A&A...479L...9C}, who stressed the important role played 
by fast rotators as ISM enrichers at low metallicities. When looking at the C 
isotope ratio as a function of the distance from the Galactic centre, 
$R_{\mathrm{GC}}$, however, we find that Models~1 and 2 \emph{without fast 
rotators} perform better. They predict an increasing trend of the ratio with 
increasing Galactocentric distance, as is indeed observed 
\citep{1994ARA&A..32..191W,2000A&A...353..349B,2002ApJ...578..211S,
2005ApJ...634.1126M}, whilst Models~3 and 4 predict a decrease in the 8--12-kpc 
Galactocentric distance range, followed by a flattening at the outermost radii. 
In order to explain the $^{12}$C/$^{13}$C ratios of both halo stars and 
molecular clouds, we conclude that fast rotating massive stars must be common 
in the early Universe, but must become rarer when the metallicity exceeds $\rm 
[Fe/H] \simeq -2$ dex. In our model, this metallicity threshold is reached only 
40\,Myr after the star formation begins, during the halo phase.

%%%%%%%%%%%%%%%%%%%%%%%%%%%%%%%%%%%%%%%%%%%%%%%%
%TWO COLUMN FIGURE

\begin{figure*}
\begin{center}
\includegraphics[width=\columnwidth]{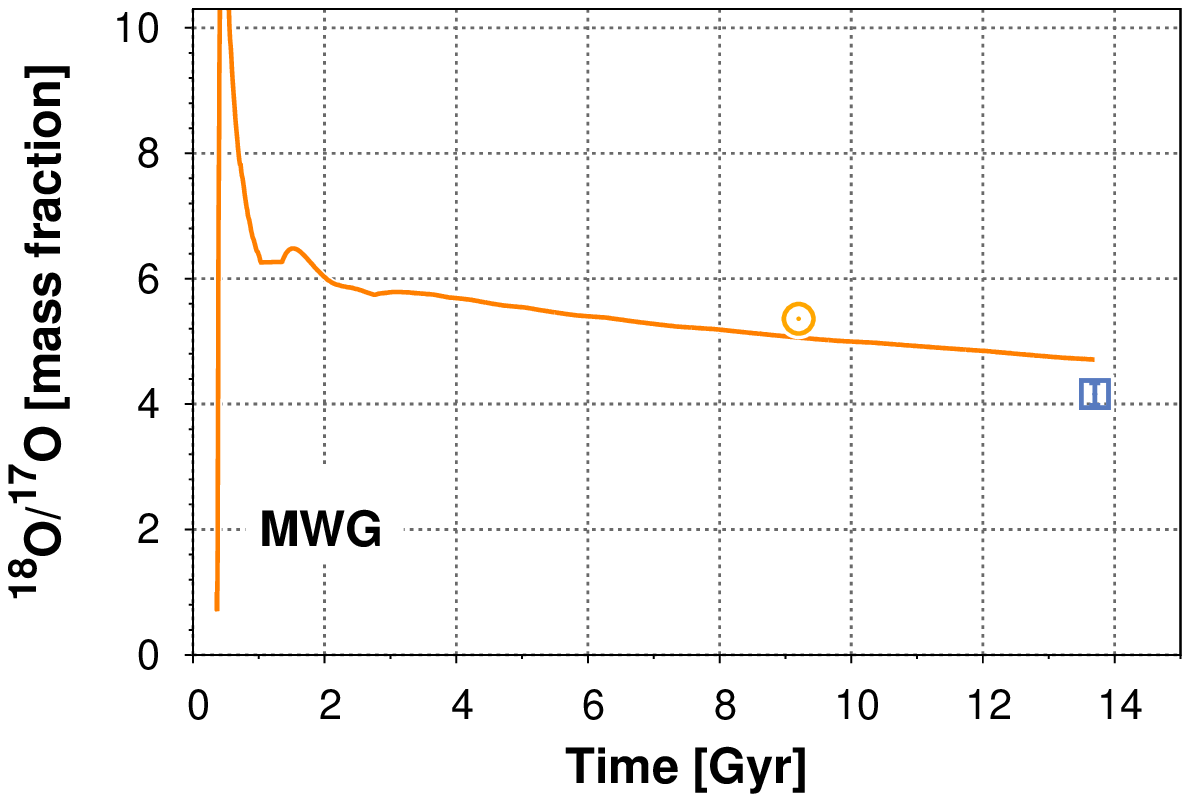}
\includegraphics[width=\columnwidth]{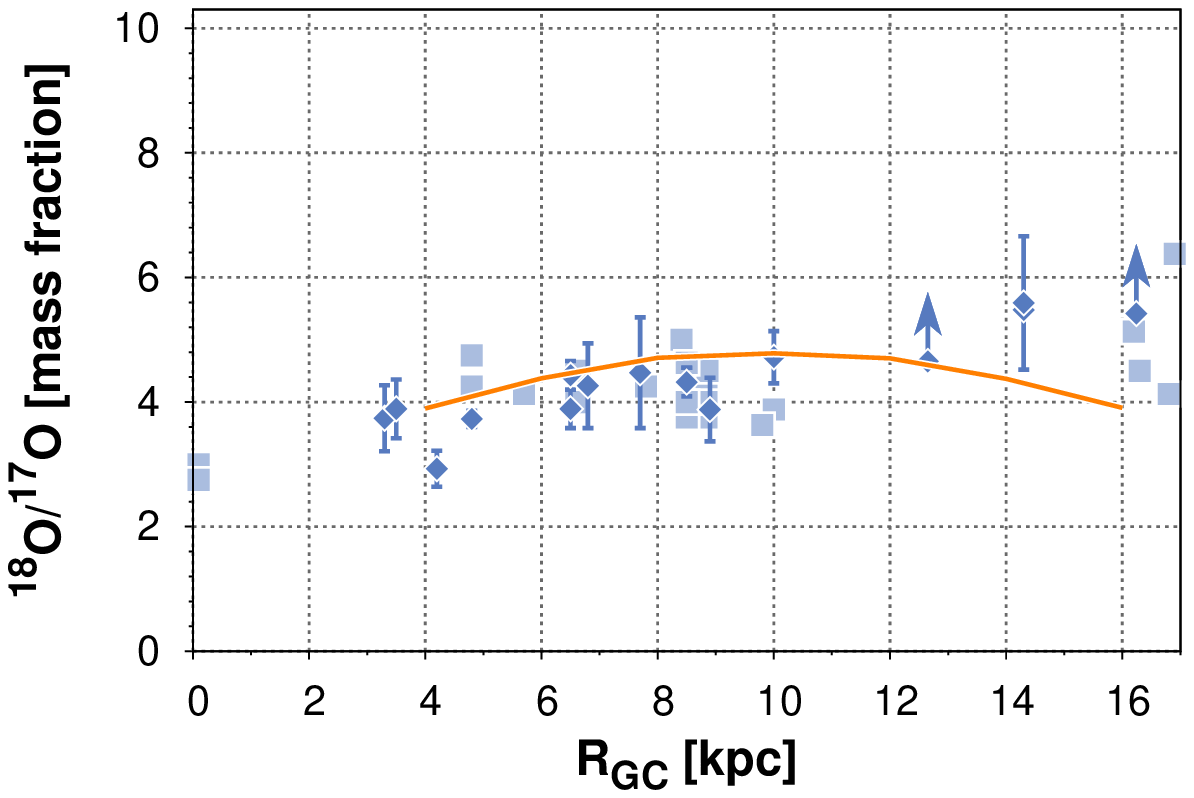}
\caption{Same as Fig.~\ref{fig:18o17oMWG}, except that theoretical predictions 
  (orange lines) are from Model~5 and include CNO production during 
  thermonuclear runaways leading to nova eruptions.}
\label{fig:18o17oNOV}
\end{center}
\end{figure*}

%%%%%%%%%%%%%%%%%%%%%%%%%%%%%%%%%%%%%%%%%%%%%%%%

Models~1 and 2 also reproduce the $^{16}$O/$^{18}$O ratios measured in the Sun 
\citep{2013ApJ...765...46A} and along the Galactic disc \citep[][the latter in 
combination with the $^{12}$C/$^{13}$C gradient of Milam et 
al.~2015]{1994ARA&A..32..191W, 2005A&A...437..957P, 2008A&A...487..237W} with 
the possible exception of the outermost regions (see Fig.~\ref{fig:16o18oMWG}, 
left and right panels, respectively). In the plots, Model~2 predictions are 
hidden behind those of Model~1 (which is expected since the two differ only in 
the treatment of super-AGB stars, while $^{16}$O and $^{18}$O are synthesised 
mostly in massive stars). The shortcomings of the models for $R_{\mathrm{GC}} 
>10$\,kpc might be due to insufficient $^{18}$O production from low-metallicity 
($\rm [Fe/H]\lesssim -1$ dex) stellar models. The $^{18}$O/$^{17}$O ratios, on 
the other hand, are severely overestimated, even when considering the 
significant contribution to $^{17}$O synthesis from super-AGB stars 
(Fig.~\ref{fig:18o17oMWG}). This may indicate the need for additional $^{17}$O 
factories, that we naturally seek in nova systems. Novae are also thought to be 
powerful $^{15}$N producers on a galactic scale.

\begin{figure*}
\begin{center}
\includegraphics[width=\columnwidth]{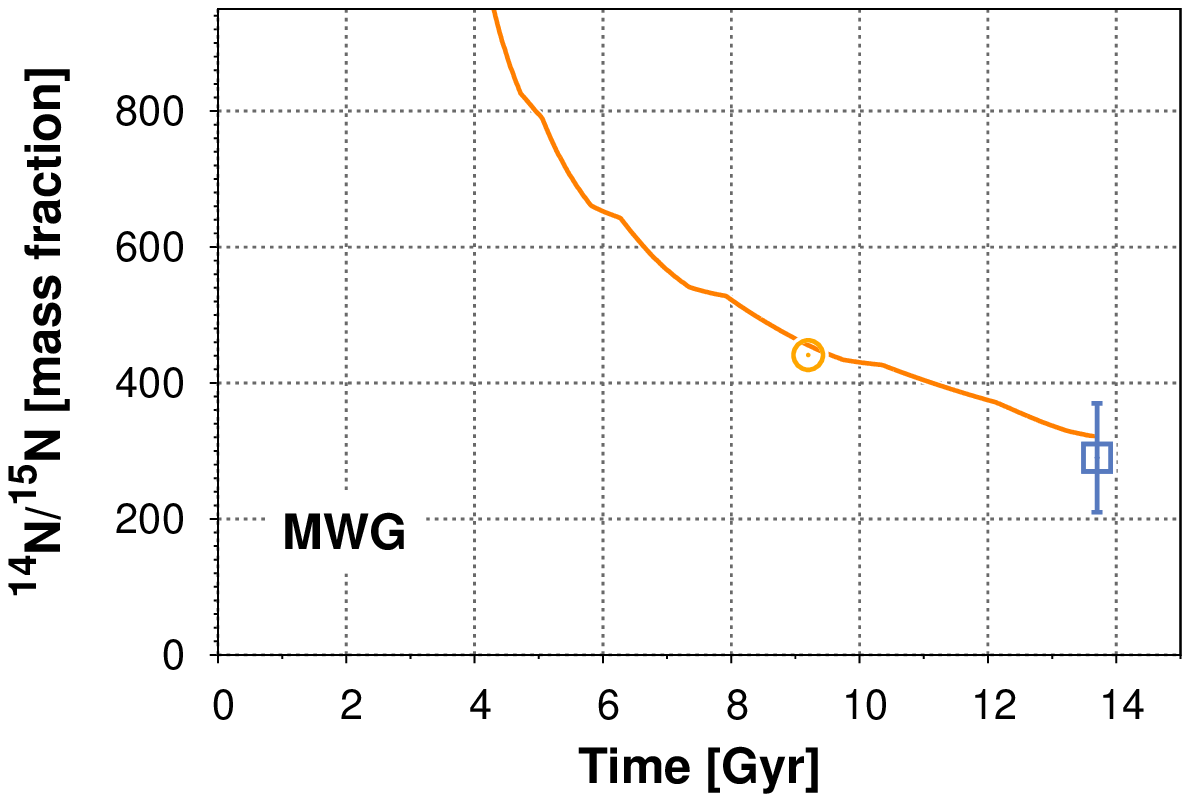}
\includegraphics[width=\columnwidth]{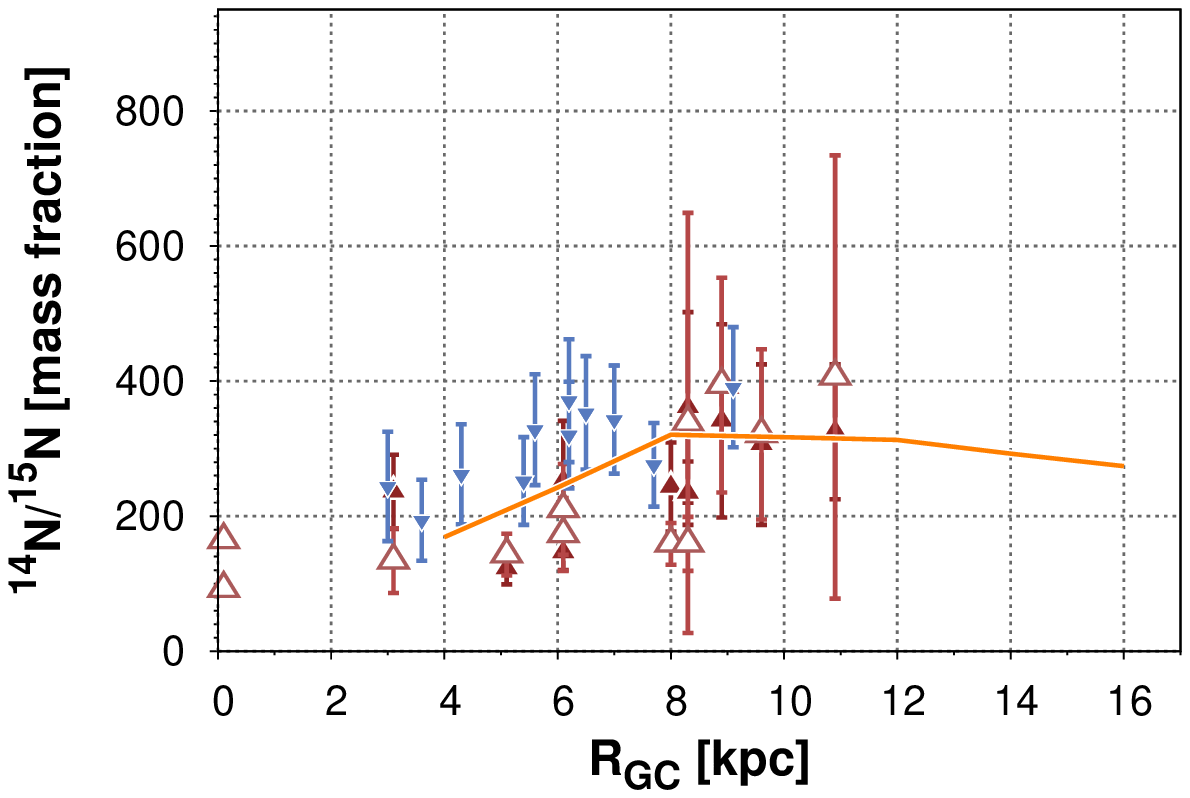}
\caption{ \emph{Left panel:} evolution of the nitrogen isotope ratio in the 
  solar vicinity predicted by Model~5 (orange line). The Sun symbol and open 
  square refer to the values of the ratio determined for the protosolar nebula 
  \citep[441$\pm$6;][]{2011Sci...332.1533M} and local ISM 
  \citep[290$\pm$40;][]{2012ApJ...744..194A}, respectively. 
  \emph{Right panel:}  nitrogen isotope ratios as a function of Galactocentric 
  distance. The solid line is the present-day gradient from Model~5, while 
  different symbols denote different observational estimates. Following 
  \citet[][their table~4]{2012ApJ...744..194A}, filled triangles represent 
  direct measurements using the hyperfine components of the CN isotopologues; 
  open triangles indicate ratios obtained by combining HN$^{13}$C/H$^{15}$NC 
  data with $^{12}$CN/$^{13}$CN measurements by \citet{2005ApJ...634.1126M}; 
  upside-down triangles show ratios derived by applying the $^{12}$CN/$^{13}$CN 
  gradient suggested by \citet{2005ApJ...634.1126M} to the HCN double 
  isotopomer data of \citet{1995A&A...295..194D}.}
\label{fig:14n15nMWG}
\end{center}
\end{figure*}

In Model~5 we have added a contribution to the CNO synthesis from novae. The 
left-hand panels of Figs~\ref{fig:18o17oNOV} and \ref{fig:14n15nMWG} show, 
respectively, the evolution of the $^{18}$O/$^{17}$O and $^{14}$N/$^{15}$N ratios 
in the solar neighbourhood; the radial behaviour of the same ratios at the 
present time are depicted in the right-hand panels. Model~5 predictions (orange 
full lines) are compared to the relevant observations. The solar 
$^{18}$O/$^{17}$O ratio \citep{2013ApJ...765...46A} can be explained by Model~5, 
which also fits the measurements across the disc (the lower-than-observed 
$^{18}$O/$^{17}$O ratios predicted by the model for $R_{\mathrm{GC}} >$~12 kpc 
likely arise from insufficient $^{18}$O production from low-metallicity massive 
stars, rather than from over-synthesis of $^{17}$O in nova outbursts --see the 
discussion in the previous paragraph). The model convincingly matches the 
nitrogen isotope ratios measured in solar wind ion samples by the {\it Genesis} 
spacecraft \citep{2011Sci...332.1533M}, in molecular clouds in the local ISM, 
and across the whole Galactic disc \citep{2012ApJ...744..194A}. The flattening 
of the theoretical $^{14}$N/$^{15}$N gradient for $R_{\mathrm{GC}} >10$\,kpc is due 
to the absence of a substantial primary $^{14}$N component in the adopted 
yields for massive stars \citep{2013ARA&A..51..457N}.  While the need for 
significant primary $^{14}$N production from low-metallicity massive stars has 
been recognized for a long time \citep[see e.g.][and references 
therein]{1986MNRAS.221..911M}, more data are needed in order to assess the 
shape of the $^{14}$N/$^{15}$N gradient at large Galactic radii, $R_{\mathrm{GC}} 
>12$\,kpc, which would place important and independent constraints on the 
amount of primary nitrogen produced by massive stars in low-metallicity 
environments.

%%%%%%%%%%%%%%%%%%%%%%%%%%%%%%%%%%%%%%%%%%%%%%%%
%TWO COLUMN FIGURE

%%%%%%%%%%%%%%%%%%%%%%%%%%%%%%%%%%%%%%%%%%%%%%%%

%%%%%%%%%%%%%%%%%%%%%%%%%%%%%%%%%%%%%%%%%%%%%%%%
%ONE COLUMN FIGURE

\begin{figure}
\begin{center}
\includegraphics[width=7.8cm]{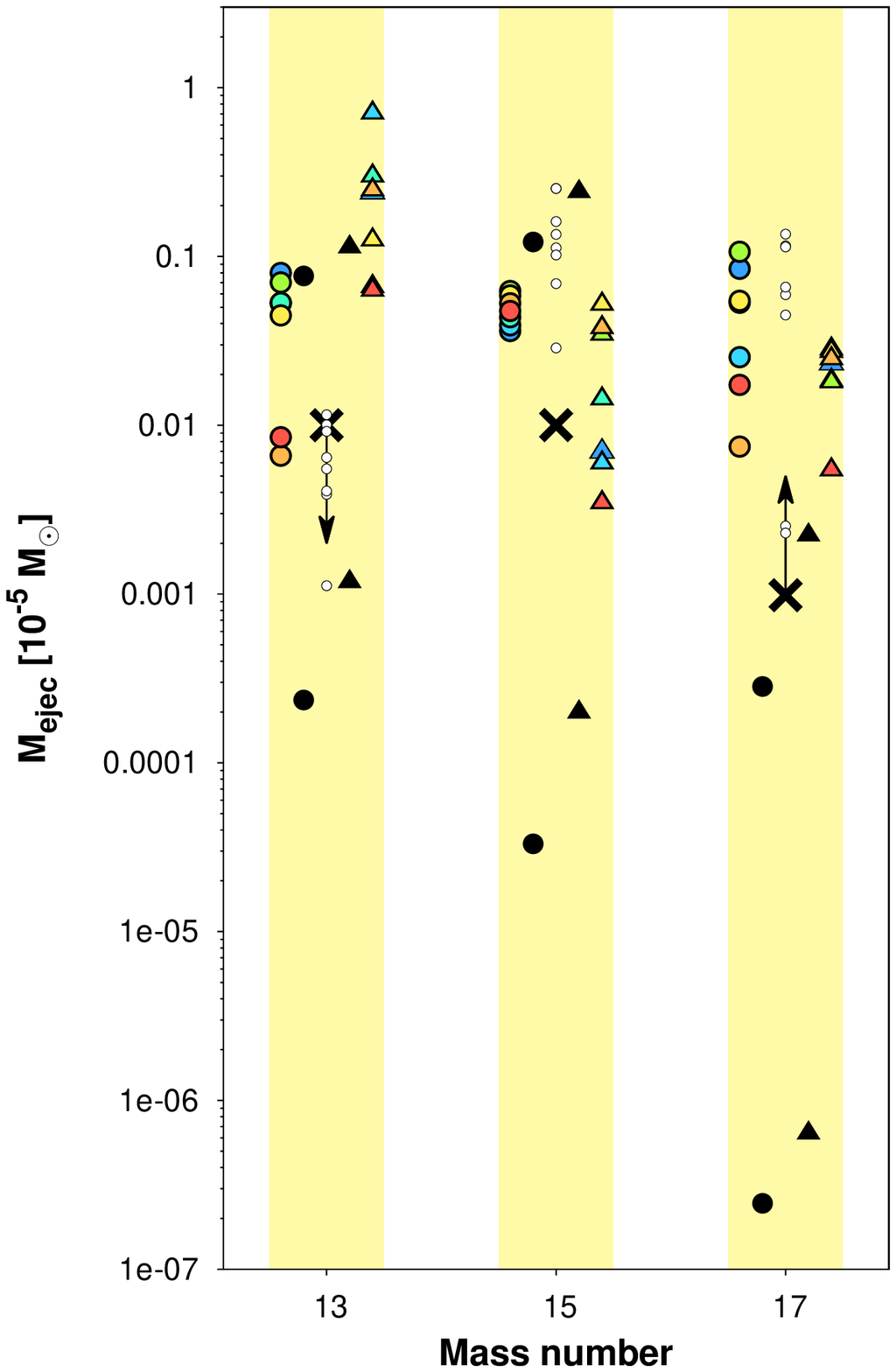}
\caption{Average masses of $^{13}$C, $^{15}$N and $^{17}$O ejected in a nova 
  outburst (units of 10$^{-5}$\,M$_\odot$) required to explain the observed 
  evolution of the CNO isotope ratios in the Milky Way in the context of our 
  chemical evolution model ($\times$ signs) contrasted with elemental yields 
  emerging from detailed hydrodynamic nova models for CO (triangles) and ONe 
  (circles) white dwarfs (coloured symbols: \citealt{1998ApJ...494..680J}; 
  black symbols: \citealt{2005ApJ...623..398Y}; small empty circles: 
  \citealt{2009ApJ...692.1532S}).}
\label{fig:novylds}
\end{center}
\end{figure}

%%%%%%%%%%%%%%%%%%%%%%%%%%%%%%%%%%%%%%%%%%%%%%%%

Only a few attempts have been made to include novae as additional CNO sources 
in Galactic chemical evolution models, even though allowing CNO production only 
from single stars fails to reproduce at least some of the CNO isotope abundance 
data available for the Milky Way \citep[e.g.][]{2003MNRAS.342..185R,
2011MNRAS.414.3231K}. This reluctance is partly driven by the ill-constrained 
parameters that describe the evolution of the close, mass-transferring binary 
systems that lead to classical nova outbursts. Over the last 45 years, however, 
hydrodynamic simulations of nova outbursts by different groups --using 
different codes and reaction rate libraries, and spanning different ranges of 
white dwarf masses and initial luminosities, mass-transfer rates from the 
main-sequence companions and mixing levels between the accreted envelope and 
the underlying white dwarf core-- have generally found ejecta enriched in 
$^{13}$C, $^{15}$N and $^{17}$O at a level that would significantly impact 
Galactic evolution \citep[see][for a recent review]{2012BASI...40..443J}. In 
Fig.~\ref{fig:novylds}, we show the masses ejected in the form of $^{13}$C, 
$^{15}$N and $^{17}$O from nova models for CO (triangles) and ONe (circles) 
white dwarfs. Coloured symbols refer to models \citep{1998ApJ...494..680J} for 
different white dwarf masses and degrees of mixing between core and envelope, 
but fixed white dwarf luminosity and mass-accretion rate. The black symbols 
represent models for different masses and mass-transfer rates, but fixed white 
dwarf temperature from \citet[][their table~6]{2005ApJ...623..398Y}. The small 
empty circles refer to simulations of 1.25-M$_{\odot}$ white dwarfs and 
1.35-M$_{\odot}$ white dwarfs using the same initial conditions but four 
different reaction-rate libraries for each mass \citep{2009ApJ...692.1532S}. It 
is evident that the mass ejected in the form of a given element in one outburst 
may vary by several orders of magnitude, depending on the specific combination 
of parameters which produces the eruption.

Following \citet[][and references therein]{1999A&A...352..117R}, in Model~5 the 
birthrate of binary systems with the characteristics necessary to give rise to 
nova eruptions is computed as a fraction $\alpha$ of the white dwarf formation 
rate\footnote{This computation takes into account a delay time of 1\,Gyr, to 
  ensure that the white dwarfs have cooled enough to lead to strong nova 
  outbursts.}. The value of $\alpha$ is fixed so as to have a current Galactic 
nova outburst rate of 20 yr$^{-1}$, consistent with the observed rate 
(20--34 yr$^{-1}$; \citealt{1994A&A...286..786D}).
%\citealt{1997ApJ...487..226S}). 
Computation of the theoretical nova outburst rate involves some knowledge of 
the average number of outbursts that a typical nova system is expected to 
experience; we take this number to be $10^4$ \citep{1978MNRAS.183..515B}. The 
average masses ejected in the form of $^{13}$C, $^{15}$N and $^{17}$O in a single 
eruption are fixed by the requirement of reproducing the observations of the 
CNO isotope ratios in the Milky Way. We find that the ejection of 
$10^{-7}$\,M$_{\odot}$ of $^{15}$N and $10^{-8}$\,M$_{\odot}$ of $^{17}$O in a single 
outburst leads to agreement between model and data. However, if current 
$^{18}$O yields from massive stars are underestimated, the latter quantity 
should be regarded as a lower limit. In fact, in this case novae should eject 
even higher amounts of $^{17}$O to compensate for the increased $^{18}$O 
production from massive stars. Now regarding $^{13}$C, we can only set an upper 
limit of $10^{-7}$\,M$_{\odot}$ per event, above which the predictions of Model~5 
deviate significantly from measurements of $^{12}$C/$^{13}$C ratios in the 
Galaxy. Our empirically determined nova yields are shown as $\times$ signs in 
Fig.~\ref{fig:novylds}.

\subsection{Other galaxies}
\label{sec:EXTres}

%%%%%%%%%%%%%%%%%%%%%%%%%%%%%%%%%%%%%%%%%%%%%%%%
%TWO COLUMNS FIGURE

\begin{figure*}
\begin{tabular}{cc}
\vspace{-0.5cm}
\includegraphics[width=\columnwidth]{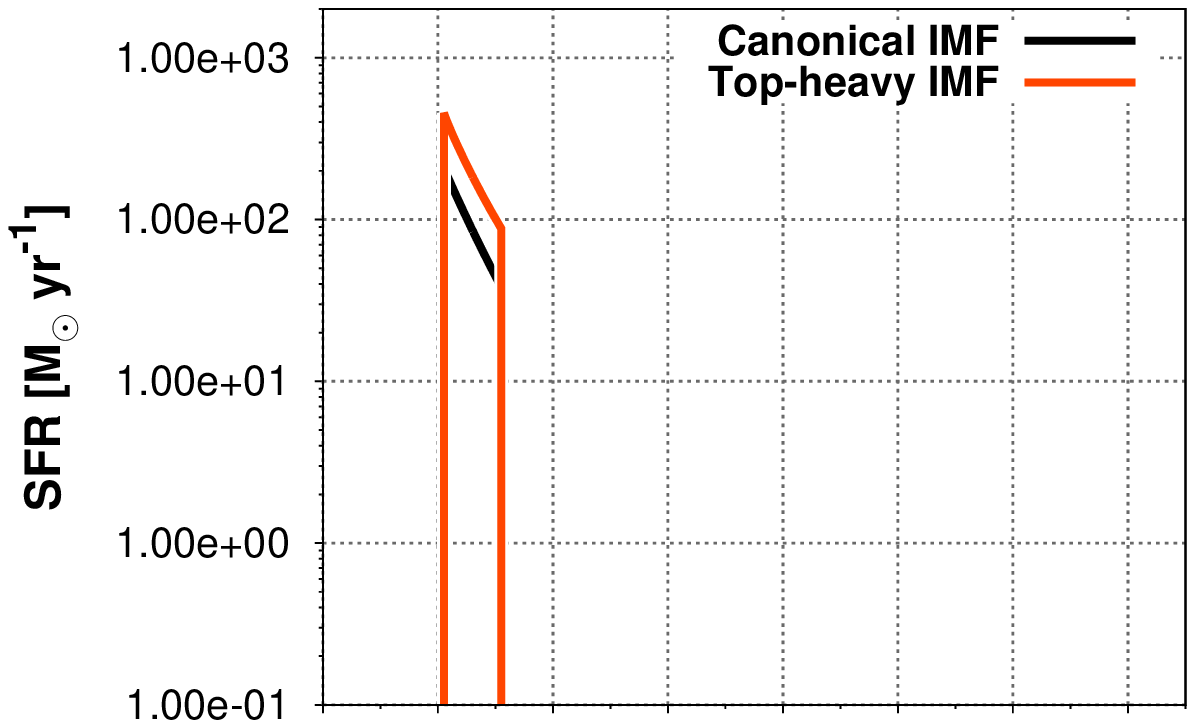} &
\includegraphics[width=\columnwidth]{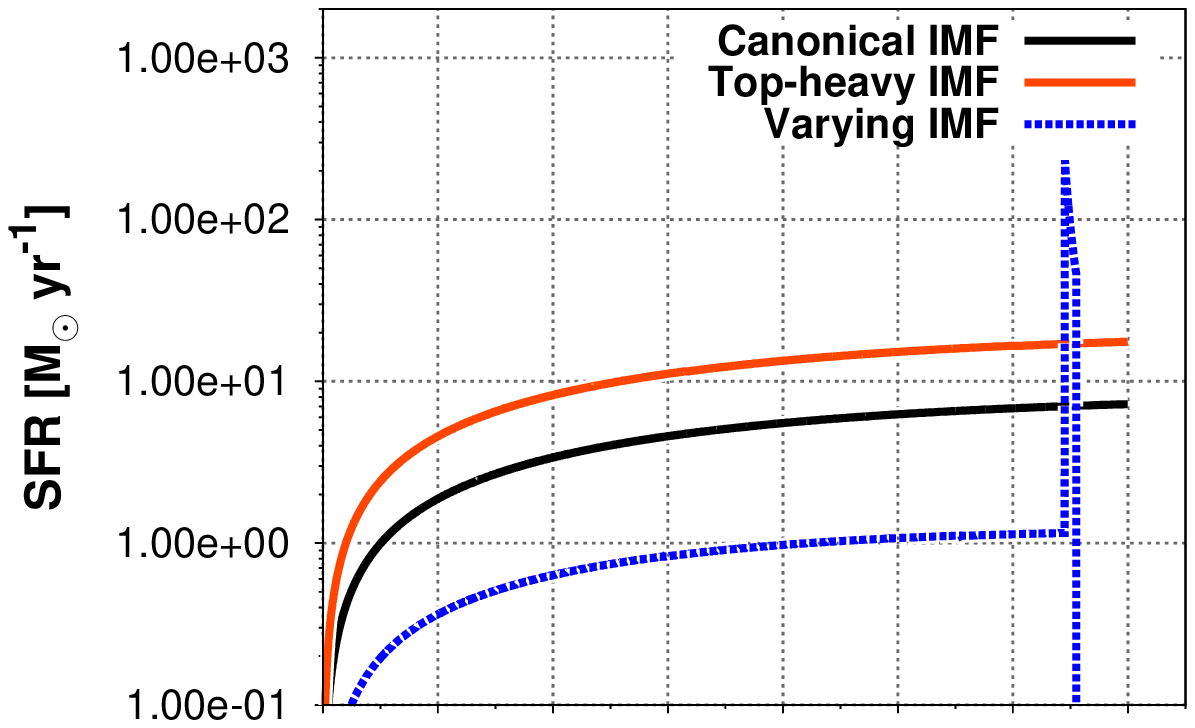} \\
\vspace{-0.5cm}
\includegraphics[width=\columnwidth]{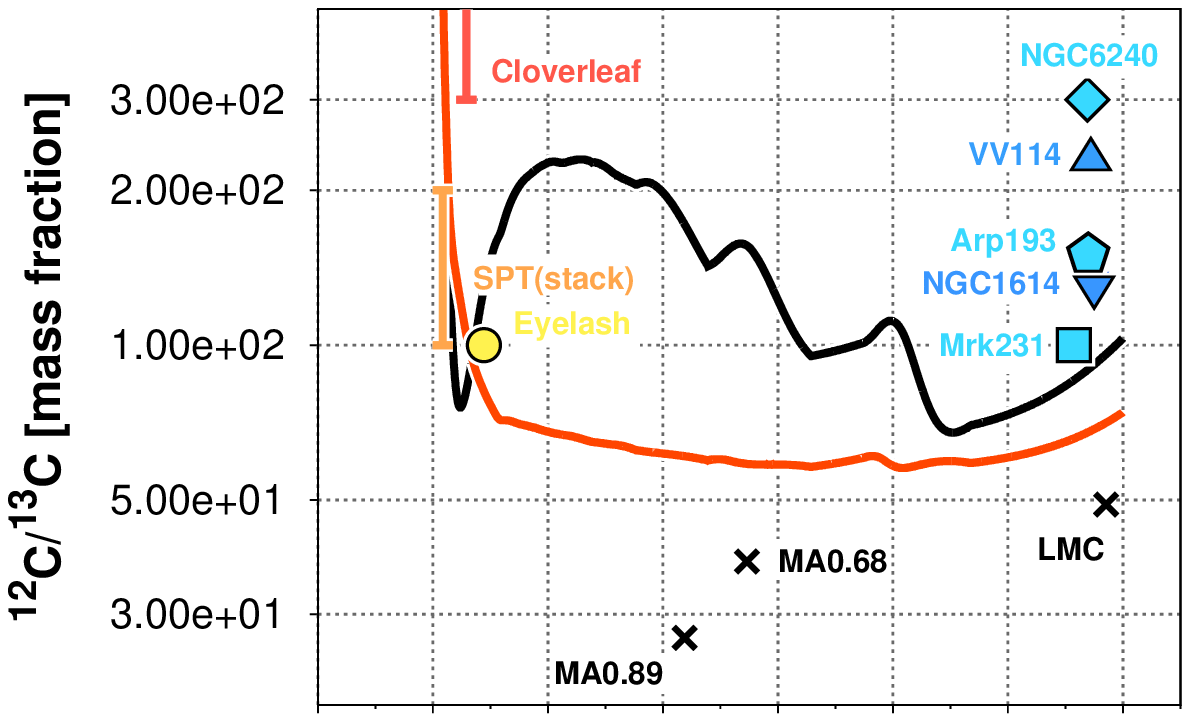} &
\includegraphics[width=\columnwidth]{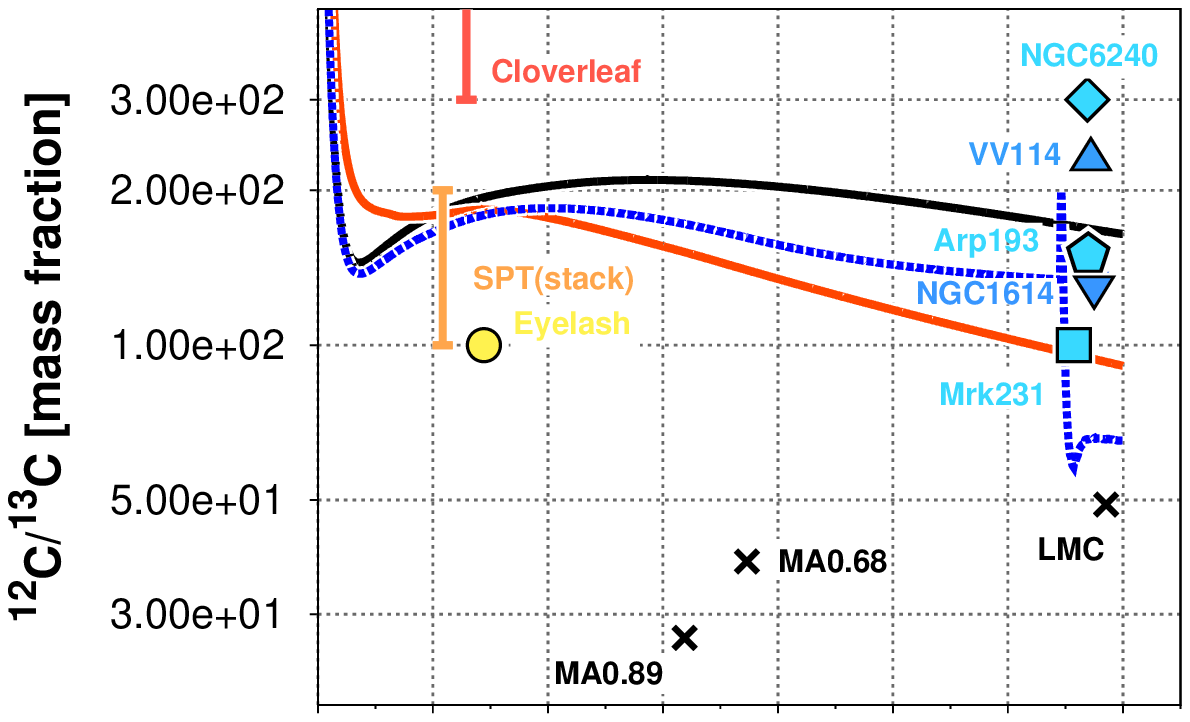} \\
\includegraphics[width=\columnwidth]{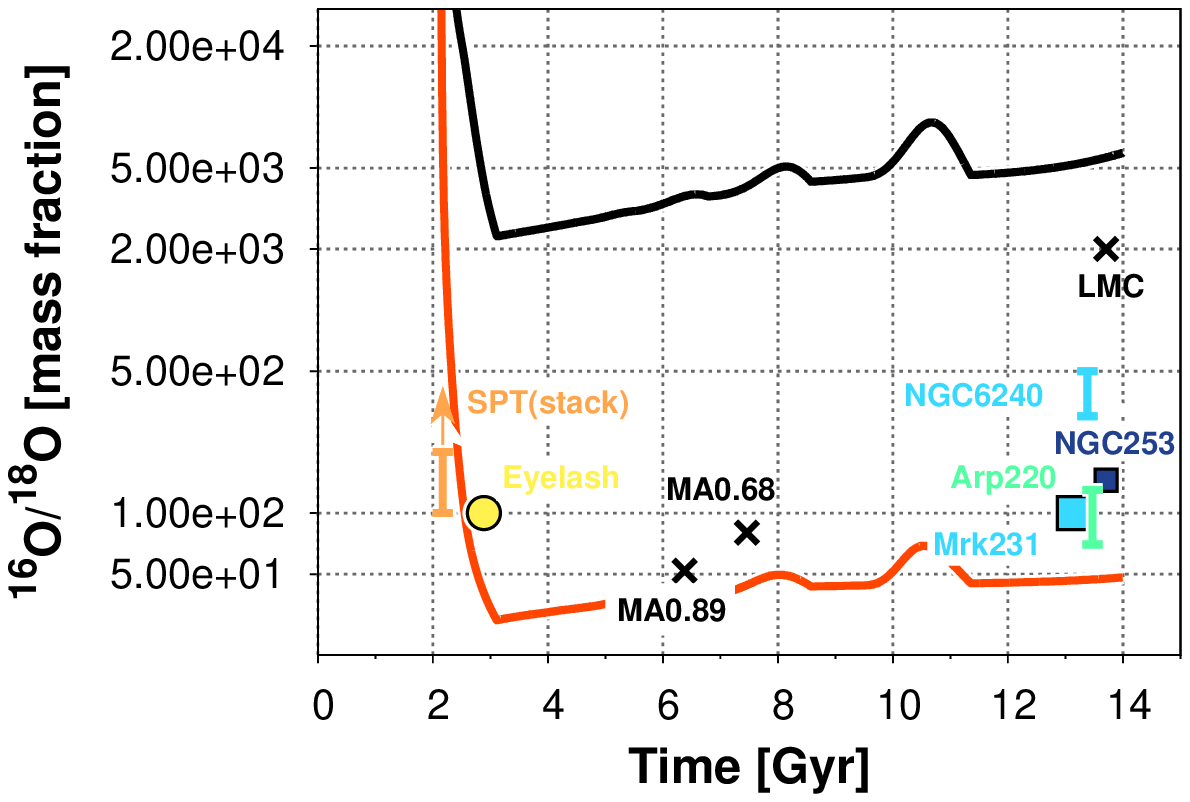} &
\includegraphics[width=\columnwidth]{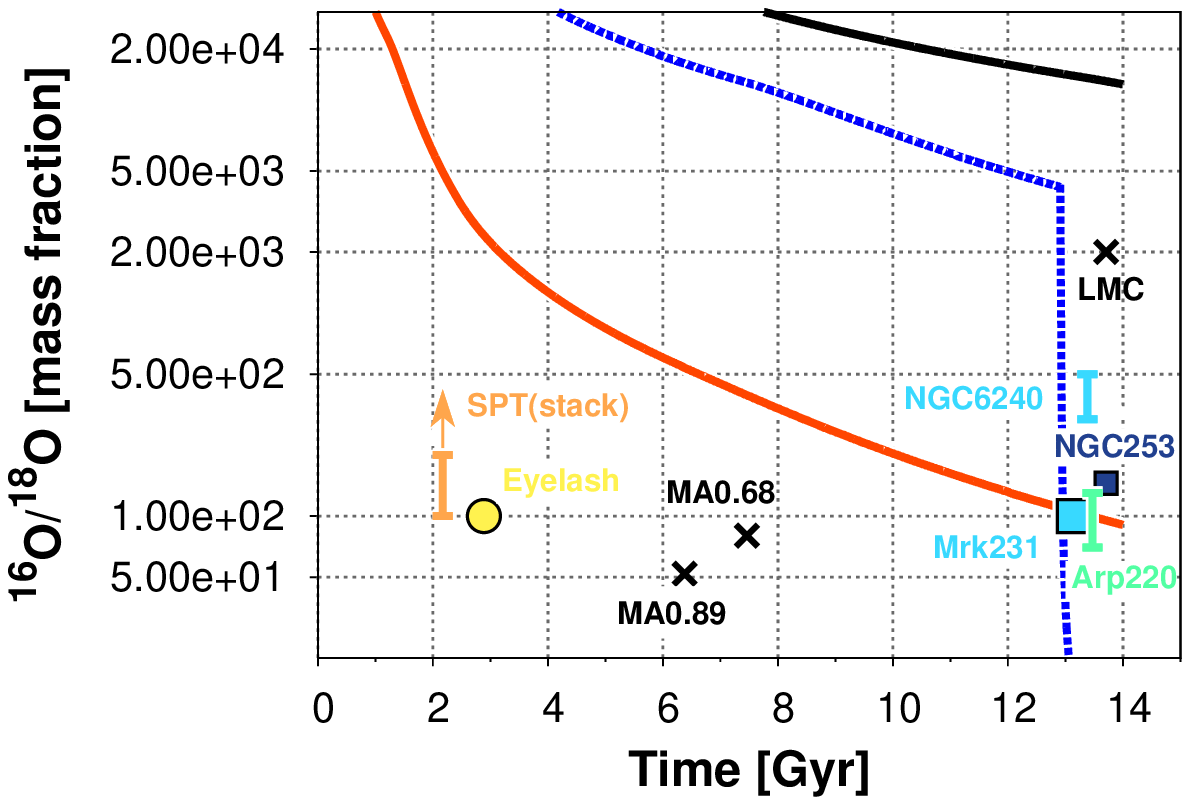}
\end{tabular}
\caption{From top to bottom: star-formation histories, evolution of the 
  $^{12}$C/$^{13}$C ratio, and evolution of the $^{16}$O/$^{18}$O ratio in the ISM 
  of galaxies resembling either high-redshift SMGs \emph{(left panels)} or 
  galaxies evolving quietly on a Hubble time \emph{(right panels)}, apart from 
  one object which experiences a late starburst (blue dashed lines).  Different 
  IMFs are considered, as labeled. The data (symbols and bars) are taken from 
  Table~\ref{tab:ext} and colour-coded according to the level of star formation 
  (see text).}
\label{fig:ext}
\end{figure*}

%%%%%%%%%%%%%%%%%%%%%%%%%%%%%%%%%%%%%%%%%%%%%%%%

In this section, we deal with the interpretation of $^{12}$C/$^{13}$C and 
$^{16}$O/$^{18}$O ratios in other galaxies found using isotopologue line 
intensities available from the literature; we do not consider objects for which 
only lower limits are provided (see \S\ref{sec:EXTdata} and 
Table~\ref{tab:ext}). In particular, we are interested in starburst galaxies, 
where the conditions of star formation differ from those typical in the local 
Universe (Papadopoulos 2010), and which may result in a top-heavy (flatter) 
IMF. As we will see, varying the IMF dramatically affects our predictions for 
CNO isotope evolution.

Already, 25 years ago, \cite{1992IAUS..149..255F} and 
\cite{1992ApJ...398...69W} speculated that IMF variations could be responsible 
for the $\alpha$-element overabundance inferred from measurements of Fe and Mg 
indices in their sample of elliptical galaxies: an IMF favouring more massive 
stars in more massive galaxies naturally leads to an overproduction of Mg 
(which is produced mostly by massive stars on short timescales) with respect to 
Fe (produced mostly by type Ia SNe with low-mass progenitors on longer 
timescales) in larger systems, as observed \citep[see 
also][]{1987A&A...185...51M}. However, they also stressed that differences in 
star-formation timescales and/or selective galactic winds could act in the same 
way, thereby leaving the issue of IMF variations unsettled \citep[see 
also][]{1994A&A...288...57M}. Nowadays, there is compelling observational 
evidence against the notion of a universal IMF \citep[see e.g. the review 
in][]{2013MNRAS.436.3309W}.

On the theoretical side, it has been suggested that when the SFR on a galactic 
scale exceeds 10\,M$_\odot$\,yr$^{-1}$, stars form in more massive and denser 
clusters, characterised by IMFs flatter than the canonical one; consequently, 
the galaxy-wide IMF also becomes top heavy \citep{2004MNRAS.350.1503W,
2013MNRAS.436.3309W,2012MNRAS.422.2246M}. Such IMF variations lead to 
significant effects on many galaxy properties, ranging from the inferred 
stellar masses and SFRs \citep[e.g.][]{2016MNRAS.462.2832C} to the detailed 
chemical composition of the galaxy constituents 
\citep[e.g.][]{2009A&A...499..711R}.

Being the ratios of primary to secondary elements, the $^{12}$C/$^{13}$C and 
$^{16}$O/$^{18}$O isotope ratios are expected to show a pronounced dependence on 
the IMF slope\footnote{$^{13}$C has both a primary and a secondary 
nucleosynthetic component.}. In the middle and lower panels of 
Fig.~\ref{fig:ext} we show, respectively, the evolution of the $^{12}$C/$^{13}$C 
and $^{16}$O/$^{18}$O ratios predicted by our models, for different assumptions 
about the mass assembly and star-formation history (see \S\ref{sec:mod} and 
Fig.~\ref{fig:ext}, upper panels), as well as the stellar IMF. The left panels 
of Fig.~\ref{fig:ext} refer to template galaxies that suffer an intense burst 
of star formation at high redshift, then evolve passively thereafter, whereas 
the right panels show the predictions for systems that experience secular 
evolution, followed by a late starburst in one case. In all panels, the black 
curves refer to models adopting a canonical IMF \citep[][with $x = 1.7$ for the 
high masses]{2002ASPC..285...86K}, while the red ones are for models with a 
top-heavy IMF \citep[][with $x = 0.95$ in the high-mass 
regime]{2007A&A...467..123B}. The blue dashed lines (right panels only) refer 
to a template galaxy that forms stars according to a canonical IMF during most 
of its lifetime, with a sudden burst of star formation initiated later, 
inducing a top-heavy IMF during the burst. The nucleosynthesis prescriptions 
are the same as in Model~1 for the Galaxy, benchmarked by its good reproduction 
of the $^{12}$C/$^{13}$C and $^{16}$O/$^{18}$O data for the Milky Way. The data 
for the galaxies listed in Table~\ref{tab:ext} (detections only) are displayed 
as symbols and bars, and are colour-coded according to the intensity of the 
star formation on a rainbow scale where red and blue stand for higher and lower 
rates, respectively.

The models on the left ($\mathscr{M}_{\mathrm{DM}} = 10^{13}$\,M$_\odot$) convert 
gas into stars at high redshift ($z \simeq 2$--3), in relatively brief 
($\Delta t_{\mathrm{burst}} =1$\,Gyr) but extremely powerful starbursts ($\rm SFR 
\simeq 100$--450\,M$_\odot$ yr$^{-1}$, for the top-heavy IMF; $\rm SFR\simeq 
40$--200\,M$_\odot$ yr$^{-1}$, for the canonical IMF). The differences in the SFR 
are dictated by the requirement that, when the star-formation activity ceases, 
a stellar mass of $\mathscr{M}_\star \simeq 2\times 10^{11}$\,M$_\odot$ must be in 
place, independently of the choice of the IMF \citep[see figure~2 in][for a 
relation between the stellar mass $\mathscr{M}_\star$ and the halo mass 
$\mathscr{M}_{\mathrm{DM}}$]{2015A&A...576L...7D}.  The modeled galaxies are 
expected to show up as SMGs at high redshifts; indeed, the theoretical 
$^{12}$C/$^{13}$C and $^{16}$O/$^{18}$O ratios compare well with observational 
estimates of these ratios in high-redshift dusty, star-forming galaxies 
\citep{2010A&A...516A.111H,2013MNRAS.436.2793D,2014ApJ...785..149S}.

In the starburst scenario we see a large variation of the  $^{12}$C/$^{13}$C 
abundance ratio for the regular IMF, with this ratio remaining significantly 
higher than for the top-heavy IMF over a period of $\sim$ 3--10~Gyr. This is 
caused by the later release of large amounts of $^{13}$C by intermediate-mass 
stars (4~$\le M/$M$_\odot \le$~7), {\it} and the much lower numbers of such 
stars in the case of a top-heavy IMF, where most of the $^{13}$C injected into 
a galaxy's ISM occurs soon, from fast-evolving high-mass stars. This 
demonstrates the need for good chemical evolution models that avoid 
instantaneous element releases, and incorporate the stellar-physics driven 
timescales for the release of the various isotopes.

Nevertheless the global $^{12}$C/$^{13}$C abundance ratio does settle to very 
similar values for both types of IMF after a period of $\sim 11$\,Gyr. Here it 
is worth emphasising that while the final anticipated global $^{12}$C/$^{13}$C 
ratio seems rather insensitive to the underlying stellar IMF, values of 
$^{16}$O/$^{18}$O as low as 100 found in the `Cosmic Eyelash' 
\citep{2010Natur.464..733S, 2013MNRAS.436.2793D} and, possibly, in the stacked 
SMG spectrum in \citet{2014ApJ...785..149S}, \emph{are reached only by an IMF 
skewed towards massive stars} (Fig.~\ref{fig:ext}, lower left panel). This 
conclusion is invariant of the assumed star-formation history; in particular, 
we tested the case of a constant or increasing SFR with time, as well as a 
regime consisting of successive bursts separated by quiescent periods. 

According to our computations, values of $^{12}$C/$^{13}$C as high as 300 --or 
even higher measured toward the Cloverleaf quasar \citep{2010A&A...516A.111H}-- 
indicate that the host galaxy is caught in a very early evolutionary phase, no 
later that $\sim 150$\,Myr from the beginning of the star-formation episode; 
concordant with this dating, we predict that measurements of the 
globally-averaged $^{16}$O/$^{18}$O ratio in the Cloverleaf should yield values 
in excess of 2,000.
 
In the right-hand panels of Fig.~\ref{fig:ext} we show the predictions of our 
models for smaller galaxies ($\mathscr{M}_{\mathrm{DM}} \simeq 10^{11}$--$10^{12}$\,
M$_\odot$, $\mathscr{M}_\star \simeq1$--$5\times 10^{10}$\,M$_\odot$) that 
experience a quieter, continuous star formation; in the case of the model for 
the lowest mass, the evolution ends after a short ($\Delta t_{\mathrm{burst}} = 
200$\,Myr), vigorous starburst ($\rm SFR\sim 100$\,M$_\odot$ yr$^{-1}$). As we 
already found for the high-redshift systems, it is immediately apparent that 
the $^{12}$C/$^{13}$C and $^{16}$O/$^{18}$O isotopic ratios for the low-redshift 
starbursts can be reproduced \emph{simultaneously} only by assuming a top-heavy 
IMF. A canonical IMF leads to $^{16}$O/$^{18}$O ratios that are higher than 
observed (Fig.~\ref{fig:ext}, lower right panel, black solid line). However, if 
the secular evolution --in which the stellar masses are distributed according 
to a canonical IMF-- is followed by a burst of star formation with a flatter 
IMF, the $^{16}$O/$^{18}$O ratio tends to values in agreement with those 
observed (Fig.~\ref{fig:ext}, lower right panel, blue dashed line). In the 
local Universe such a sequence of galaxy evolution events could be mirrored by 
two gas-rich spirals evolving in isolation, before they merge inducing a strong 
merger/starburst late in their histories. This could indeed be the story behind 
local gas-rich starbursts, which makes the comparison of their global 
isotopologue ratios with the predictions of our `starburst model' appropriate.

Finally, while a detailed analysis of the data available for local starbursts 
is beyond the scope of the present paper, it is worth noticing that our 
`starburst model' predicts $^{12}$C/$^{13}$C and $^{16}$O/$^{18}$O ratios that 
span almost the full range of values covered by observations of local systems. 
Nevertheless, regardless of the details of the underlying star-formation 
scenario, a top-heavy IMF seems necessary for building some part of their 
eventual stellar mass, with the low $^{16}$O/$^{18}$O abundance ratios proving 
decisive. The fact that such ratios have been implied for many years, even for 
isolated (but vigorously star-forming) spiral galaxies 
\citep[e.g.,][]{1996ApJ...465..173P}, but nevertheless could not be attributed 
uniquely to a top-heavy IMF, underscores the value of good CNO evolutionary 
models in setting such important constraints.

\section{Conclusions}
\label{sec:end}

In this paper, we use state-of-the-art chemical evolution models to track the 
evolution of CNO isotopes in the ISM of galaxies. We take advantage of recently 
published stellar yields, as well as new isotopologue molecular line data for 
the Milky Way and many other galaxies. First, we use a multi-zone model for the 
Milky Way to re-assess the relative roles that AGB stars, massive stars, and 
novae have in the production of the CNO elements, focusing on the rare 
isotopes. We find that:

\begin{enumerate}
\item In order to reproduce the $^{12}$C/$^{13}$C ratios observed in both nearby 
  halo stars and molecular clouds spanning a large range of Galactocentric 
  distances, fast-rotating massive stars must be common in the early Universe, 
  and become rarer when a metallicity threshold of $\rm [Fe/H]\simeq -2$\,dex 
  is reached; this happens very early on in our model, about 40\,Myr after the 
  beginning of the star formation in the halo phase.
\item Super-AGB stars synthesise significant amounts of $^{17}$O; yet, in order 
  to bring the theoretical predictions for the $^{18}$O/$^{17}$O ratio into 
  agreement with the relevant Galactic data, one also needs to consider a 
  contribution to $^{17}$O synthesis from novae.
\item Novae are basic producers of $^{15}$N; taking their contribution into 
  account, we are able to reproduce both the declining trend of 
  $^{14}$N/$^{15}$N in the solar neighbourhood suggested by recent 
  determinations of this ratio for the protosolar nebula 
  \citep{2011Sci...332.1533M} and the $^{14}$N/$^{15}$N Galactic gradient 
  recently revised by \cite{2012ApJ...744..194A}. Hydrogen ingestion into the 
  helium shell of core-collapse supernovae has been suggested as an additional 
  source of the rare $^{15}$N isotope by \cite{2015ApJ...808L..43P} who, 
  however, do not provide a full grid of yields for use in chemical evolution 
  models. This prevents us from testing their interesting proposal by means of 
  our models.
\item By assuming yields from \cite{2010MNRAS.403.1413K} for low- and 
  intermediate-mass stars, from \cite{2013ARA&A..51..457N} for massive stars, 
  and average ejected masses of $M_{\rm ejec}^{^{13}{\rm C}} \le 10^{-7}$\,M$_\odot$, 
  $M_{\rm ejec}^{^{15}{\rm N}}= 10^{-7}$\,M$_\odot$ and $M_{\rm ejec}^{^{17}{\rm O}}\ge 
  10^{-8}$\,M$_\odot$ per nova outburst, we find excellent agreement between most 
  of our model predictions and the Milky Way CNO data; the model could be 
  further constrained and improved when new data on the Galactic gradient at 
  large radii become available.
\end{enumerate}

Furthermore, we show that as far as galaxy-averaged abundance ratios are 
concerned, neither FUV-driven selective photodissociation nor chemical 
fractionation can seriously `skew' isotopologue ratios away from isotopic ones. 
Then after selecting the best set of yields, we use single-zone models for 
other galaxies to investigate \emph{qualitatively} the dependence of the 
$^{12}$C/$^{13}$C and $^{16}$O/$^{18}$O ratios on the star-formation history and 
stellar IMF. By comparing the model predictions to observations of isotopologue 
(and thus isotopic) abundance ratios for the molecular gas reservoirs of 
high- and low-redshift starbursts, we conclude that:

\begin{enumerate}
\item In order to explain \emph{simultaneously} the measurements of 
  $^{12}$C/$^{13}$C and $^{16}$O/$^{18}$O ratios in starbursts, an IMF skewed 
  towards high stellar masses is needed; this requirement is driven by the 
  higher-than-observed $^{16}$O/$^{18}$O ratios obtained with a canonical IMF, 
  whilst the final $^{12}$C/$^{13}$C ratio is not dramatically affected by the 
  choice of the IMF.
\item The high $^{12}$C/$^{13}$C ratios observed in some systems imply that 
  either the galaxy is caught in its very early phases of evolution or, 
  alternatively, that it is being `rejuvenated' by a starburst that has erased 
  the memory of the preceding evolution and reset the $^{12}$C/$^{13}$C ratio to 
  a high value; this, however, \emph{is possible only if the IMF becomes 
    top-heavy during the starburst}.
\item For the host of the Cloverleaf quasar, at $z \simeq 2.5$, given the 
  observational estimate of 300--10,000 for the $^{12}$C/$^{13}$C ratio, we 
  predict that a value of $^{16}$O/$^{18}$O in excess of 2,000 should be found.
\item Changing the details of the CNO isotope evolution, or the assumptions 
  about the histories of mass assembly and star formation does not affect the 
  main conclusion that a top-heavy IMF is needed in starbursts.
\end{enumerate}

In closing, we re-iterate how important these issues are for galaxy 
evolution, with ALMA now able to provide the data necessary to re-examine them, 
in the context of the concurrent CNO isotope evolution models, which can then 
give us the ages of starburst events and the prevailing stellar IMFs across 
cosmic time.

\section*{Acknowledgements}

This research was supported by the Munich Institute for Astro- and Particle
Physics (MIAPP) of the DFG cluster of excellence \emph{``Origin and Structure
of the Universe''}. This work also benefited from the International Space
Science Institute (ISSI) in Bern, thanks to the funding of the team \emph{``The
Formation and Evolution of the Galactic Halo''} (PI D.~Romano). PPP is
supported by an Ernest Rutherford Fellowship.  RJI and ZYZ acknowledge support
from ERC in the form of the Advanced Investigator Programme, 321302, COSMICISM.

\bibliographystyle{mnras}
\bibliography{dromano}

     \label{lastpage}

\end{document}